\documentclass{article}

\usepackage{PRIMEarxiv}

\usepackage[utf8]{inputenc} 
\usepackage[T1]{fontenc}    
\usepackage{hyperref}       
\usepackage{url}            
\usepackage{booktabs}       
\usepackage{amsfonts}       
\usepackage{nicefrac}       
\usepackage{microtype}      
\usepackage{lipsum}
\usepackage{fancyhdr}       
\usepackage{graphicx}       
\usepackage{subcaption}
\usepackage{pifont}
\usepackage{amsmath,amssymb,amsfonts}%
\usepackage{amsthm}%
\usepackage{tabularx}
\graphicspath{{media/}}     

\title{QMedShield: A Novel Quantum Chaos-based Image Encryption Scheme for Secure Medical Image Storage in the Cloud}

\author{
  Arun Amaithi Rajan \\
  \textbf{ORC ID:} $0000-0002-3019-5879$\\
  Department of Computer Science and Engineering \\
  College of Engineering Guindy, Anna University \\
  Chennai, India\\
  \texttt{22144191119@student.annauniv.edu} \\
  \And 
  Vetriselvi V \\
  \textbf{ORC ID:} $0000-0002-3832-6968$\\
  Department of Computer Science and Engineering \\
  College of Engineering Guindy, Anna University \\
  Chennai, India\\
  \texttt{vetri@annauniv.edu}} 

\begin{document}
\maketitle

\begin{abstract}
In the age of digital technology, medical images play a crucial role in the healthcare industry which aids surgeons in making precise decisions and reducing the diagnosis time. However, the storage of large amounts of these images in third-party cloud services raises privacy and security concerns. There are a lot of classical security mechanisms to protect them. Although, the advent of quantum computing entails the development of quantum-based encryption models for healthcare. Hence, we introduce a novel quantum chaos-based encryption scheme for medical images in this article. The model comprises bit-plane scrambling, quantum logistic map, quantum operations in the diffusion phase and hybrid chaotic map, DNA encoding, and computations in the confusion phase to transform the plain medical image into a cipher medical image. The proposed scheme has been evaluated using multiple statistical measures and validated against more attacks such as differential attacks with three different medical datasets. Hence the introduced encryption model has proved to be attack-resistant and robust than other existing image encryption schemes, ensuring the secure storage of medical images in cloud environments
\end{abstract}


\section{Introduction}\label{sec1}

In the contemporary landscape, digital images serve as important tools across several domains, facilitating communication, documentation, analysis, and creative expression.  Recent statistics indicate that 1.81 trillion pictures are captured each year, which is anticipated to rise by 10–14\% each year just from mobile phones. By 2030, experts predict that the we might have 30 trillion images \cite{broz-2023}. From the numbers, It is clear that images play a significant role in conveying information and enhancing understanding.\\

In the realm of healthcare, images have become indispensable nowadays. Medical images contain more intricate and important information about the patient. By using advanced imaging technologies, internal structures, and abnormalities can be visualized by healthcare professionals, and identify the progression of diseases with better clarity and precision. They are helping the physicians to make timely decisions \cite{Mercaldo2023ObjectLocalization}. As these highly sensitive medical images increase exponentially, are being stored in third-party cloud servers which are prone to cyber-attacks \cite{Lovrencic2023Multi-cloudSecurity}. Given the paramount importance of security and privacy in medical image storage, it is imperative that all medical information be securely stored, with the responsibility for implementing necessary measures resting on the information owners. Even minor modifications to those medical images lead to erroneous diagnoses and improper information retrieval secure \cite{Liu2024SecureTechnique}. This can be achieved through medical image encryption techniques and ensures the secureness of the medical image in the cloud storage. The process of secure storage of medical images is illustrated in Figure 1. 

\begin{figure}[h!]
\centering
\includegraphics[width=0.8\textwidth]{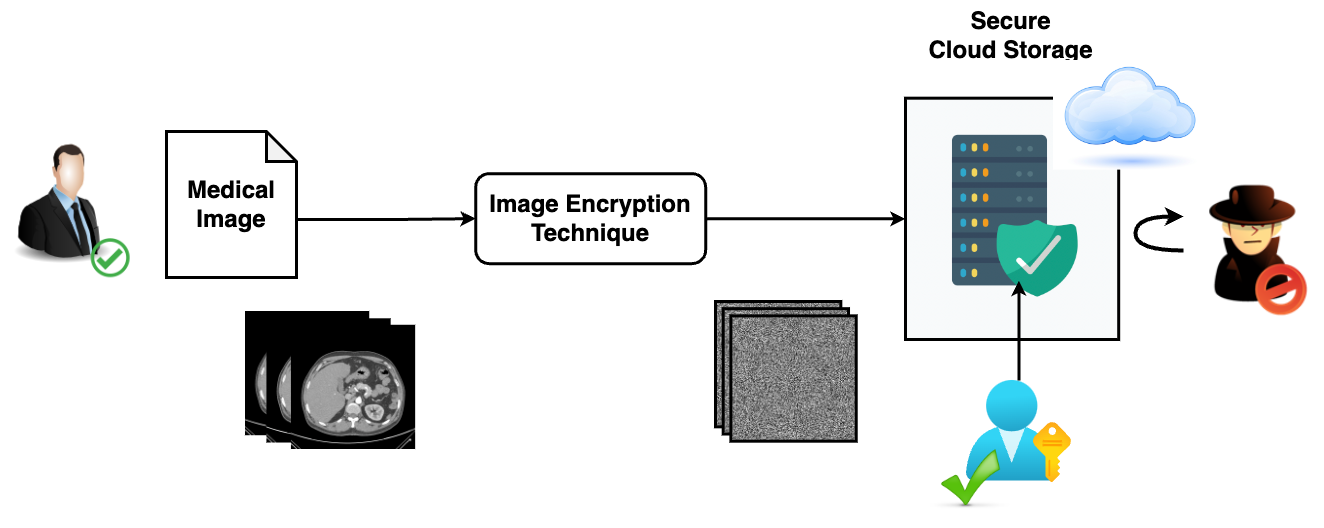}
\caption{Secure Cloud Storage} 
\label{fig1}
\end{figure}

In the insecure cloud, encrypting images emerges as the most efficient and secure technique for their storage. There exist various image encryption techniques. Kaur et al. \cite{Kaur2020ATechniques} categorized them and discussed each method's pros and cons. Most of the image encryption methods bring randomness to the image pixels by introducing chaotic maps. These maps are available in different dimensions and the generated sequences are super sensitive to control parameters and initial values, great ergodicity, and greater pseudo-randomness \cite{Wang2020AMulti-chaos}. As these chaotic sequences demonstrate randomness, scrambling can be introduced among the bit-planes to improve diffusion. Furthermore, to enable space-efficient pixel substitution, DNA encoding presents a viable solution, allowing operations to be conducted over DNA-encoded data \cite{Xu2022DNAData}. Consequently, a set of image encryption algorithms is proposed by integrating DNA computing with chaotic maps. Those techniques are effective and robust. However, From the quantum computing standpoint, these schemes are more susceptible to various attacks.  Leveraging quantum-based chaotic maps can mitigate these vulnerabilities, as quantum gates introduce randomness and variability, and enhance security against attacks. Additionally, qubit's unique properties, including entanglement, offer robustness against classical and potential quantum threats, improving the security of image encryption. Consequently, there is a growing need for a secure image encryption model using quantum chaos in digital healthcare \cite{Liu2024QuantumChaos}. \\

Hence, In this work, we introduced a new encryption technique for medical images, QMedShield, integrating quantum chaotic maps, bit plane scrambling, quantum operations, and hybrid chaotic maps, with DNA encoding techniques. The overall block diagram is depicted in Figure 2. The major contributions to the article are outlined below. 

\begin{enumerate}
    \item A New medical image encryption scheme, QMedShield is proposed for their secure storage in the cloud with quantum chaos, quantum operations, hybrid chaos, bit plane scrambling, and DNA encoding.
    \item Pixel diffusion is achieved with bit plane scrambling, classic and quantum chaotic sequences, and quantum operations such as Hadamard and CNOT.
    \item Pixel confusion has been done with hybrid chaotic map sequence and DNA encoding-based pixel substitution. 
    \item QMedShield has validated using three different medical image datasets and demonstrated that the model is secure and attack-resistant with multiple statistical experimental analyses. 
\end{enumerate}

\begin{figure}[h!]
\centering
\includegraphics[width=0.9\textwidth]{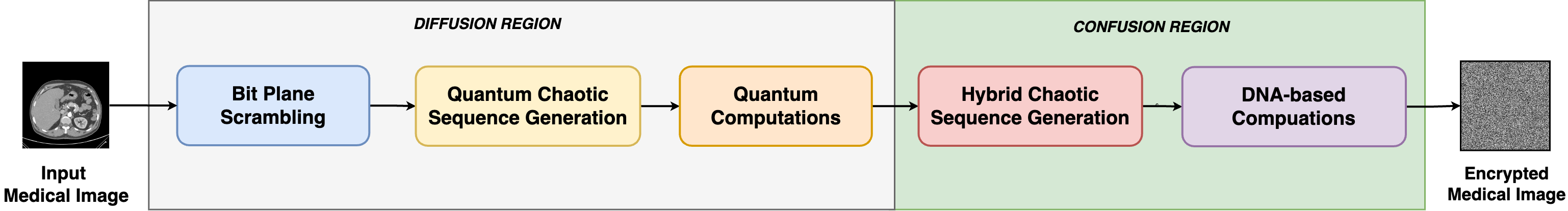}
\caption{Overall Block Diagram} 
\label{fig1}
\end{figure}

The subsequent segments of the article are organized as follows: Section 2 discusses the different image encryption schemes available. Preliminary concepts are briefed in section 3. Section 4 elaborates on the novel encryption algorithm, QMedShield. In Section 5, the security analysis of the introduced encryption model is discussed. In the end, Section 6 presents the paper's conclusion and outlines future directions.

\section{Related Work}

This section explores the existing image encryption schemes and their evolution in recent times.  Some survey articles already discuss medical image encryption from different aspects \cite{Magdy2022SecurityReview}. Through classical image encryption, sensitive visual data is protected from unauthorized access and malicious exploitation. With robust primitives like substitution-permutation networks and Feistel ciphers, classical encryption methods aim to obfuscate image content and resist cryptanalysis. Classical image encryption entails two primary domains: spatial and transform \cite{Kaur2020ATechniques}. There are increasingly interesting algorithms utilizing DNA encoding, cellular automata, metaheuristics, chaotic maps, fuzzy logic, and more in the spatial domain. Healthcare is one of the most important domains, which requires this kind of secure and confidential image storage techniques. Priyanka et al. \cite{Priyanka2022AApplications} offer a detailed exploration of encryption techniques applicable to healthcare images, along with suggestions for future research directions.\\

 In cryptography, chaotic maps offer a promising avenue for generating secure cryptographic keys by exploiting the chaotic behavior exhibited by nonlinear dynamical systems \cite{Al-Hazaimeh2019ImageKeys}. Additionally, chaotic maps serve as effective components in image encryption schemes, where they facilitate the transformation of plaintext images into ciphered forms by introducing complexity and randomness. Different chaotic maps are used such as logistic maps \cite{Balasamy2021ASVD}, 2D Arnald maps \cite{Khare2021AImages}, Chen \cite{Yousif2022ATechniques}, Lorenz \cite{Zhang2022MultipleAlgorithm}, Henon \cite{Kumar2023EnhancingTable}, and Tent maps \cite{AmaithiRajan2023SecureEncoding} are used for multiple purposes like confusion, diffusion, random key generation, scrambling, and substitution. Paul et al. \cite{Paul2022ASHA-2} developed a unique encryption scheme that integrates hyperchaotic maps and Zaslavskii map-based pixel shifting with SHA-2 algorithm. Recently Yi et al. \cite{Yi2023AnModeling} also introduced a novel image encryption mechanism that leverages classic AES and the latest Rossler hyperchaotic systems. To improve the performance, image compression-based encryption mechanisms are developed \cite{Ahmad2022Encryption-then-CompressionServices}. DNA encoding enhances the diffusion rate in image encryption schemes.  Bao et al. \cite{Bao2022ACoding} processed the DNA-encoded image with Chen chaotic map and a compressive sensing technique. Similarly, Arthi et al. \cite{Arthi20224DCommunication} devised a 4D-hyperchaotic map combined with a DNA-encoding technique and validated with medical images. Moreover, In multimedia, DNA-computing-based encryption techniques have been broadening to cloud applications \cite{Namasudra2021SecuringEnvironment}. Amaithi Rajan et al. \cite{AmaithiRajan2023SecureEncoding} designed an encryption model for the secure storage and processing of medical images in a cloud server, utilizing hyperchaotic maps, DNA encoding, and bitplane scrambling. The model demonstrates robustness and resistance against attacks, ensuring the security and privacy of sensitive image information.\\

However, all these techniques are classical cryptography-based, which are easily compromised by quantum computing. So, we are in need of a based image encryption model. Researchers are working in this area also. A framework for the chaos-based  \cite{Yan2022TowardComputers}quantum encryption of healthcare images that guarantees patient safety and anonymity was presented in an article by Abd El-Latif et al. \cite{AbdEl-Latif2017RobustImages}. Janani et al. \cite{Janani2021ARepresentation} also proposed a new technique based on quantum image representation and chaotic maps to secure medical images. \\

Based on the insights from the literature review, integrating quantum chaotic maps, quantum computations, bit-plane scrambling and DNA encoding could increase image encryption security. In line with these findings, we devised a novel image encryption model. Table 1 shows how our method varies from existing methods.

\begin{table}[h!]
\centering
\caption{Our Method Vs Existing Methods}
\begin{tabularx}{\textwidth}{X|X|X|X|X|X|X|X}
\hline
 \textbf{Ref }& \multicolumn{2}{X|}{\textbf{Chotic Map}} & \textbf{DNA Computing} & \textbf{Quantum Operations} & \textbf{Bit Plane Scrambling} & \textbf{Stat Attack Resistance} & \textbf{CP/KP Attack Resistance} \\ \hline
             &\textbf{ Classic Chaotic Map} & \textbf{Quantum Chaotic Map} &    &    &    &    &    \\ \hline
\cite{Arthi20224DCommunication}            & \ding{52}                  &                     & \ding{52} &    &    & \ding{52} & \ding{52} \\ \hline
\cite{Houshmand2024OptimizedImages} & \ding{52}                  &                     &    & \ding{52} &    & \ding{52} &    \\ \hline
\cite{AmaithiRajan2023SecureEncoding}           & \ding{52}                  &                     & \ding{52} &    & \ding{52} & \ding{52} & \ding{52} \\ \hline
\cite{Patel2024SecuredMethods}            & \ding{52}                  &                     &    & \ding{52} &    & \ding{52} &    \\ \hline
Ours            & \ding{52}                 & \ding{52}                  & \ding{52} & \ding{52} & \ding{52} & \ding{52} & \ding{52} \\ \hline
\end{tabularx}
\end{table}

\section{Preliminaries}

This section outlines the basic definitions of concepts used in the encryption model employed. It covers topics such as bit plane scrambling, different chaotic maps, basic quantum operators, and DNA encoding.

\subsection{Bit Plane Scrambling}

The bit plane of a digital image comprises an array of bits corresponding to particular bit positions in each of the binary representations of a pixel \cite{AmaithiRajan2023SecureEncoding}. For example,  in an image where each pixel is represented by 8 bits, there are 8-bit planes. As illustrated in Figure 3, bit plane scrambling rearranges these planes according to a key. The benefits of bit plane scrambling include improving image security by diffusing visual patterns, strengthening defenses against unauthorized access or manipulation, and facilitating encryption efficiency while preserving image fidelity.\\

\begin{figure}[h!]
\centering
\includegraphics[width=0.6\textwidth]{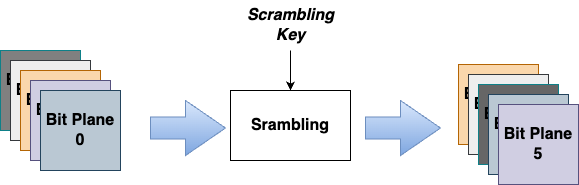}
\caption{Bit plane Scrambling} 
\label{fig1}
\end{figure}

\subsection{Chaotic Maps}

Chaotic maps explore the behaviors of dynamic systems that often express non-linear randomness. They are categorized into two types: 1D and nD systems. They are too sensitive to initial conditions \cite{Zia2022SurveyDomains}. This implies that outputs can be drastically altered by slightly changing the initial parameters. Chaotic maps can be categorized into discrete and continuous maps, both serving to generate keys for pixel diffusion in images. We employ a 2D Henon map, a hybrid logistic-sine map, and a 3D quantum logistic map in the proposed encryption algorithm.  Subsequent subsections provide a comprehensive explanation of the logic behind these chaotic maps.

\subsubsection{Henon Chaotic Map}
Henon map is a 2D quadratic chaotic map  \cite{Kumar2023EnhancingTable}. This Henon 2D chaotic map provides significant advantages over 1D chaotic maps in image encryption models. Its two-dimensional nature enables richer and more complex chaotic behavior, leading to increased security and robustness against cryptanalysis. Moreover, the Henon map's strong nonlinear behavior introduces randomness and complexity into the encryption process, improving the security of the encrypted image. Its efficient diffusion properties ensure that changes to individual pixels propagate effectively, enhancing the overall security and cryptographic strength of the encryption model. This map is defined using the equations 1 and 2 below.

\begin{equation}
    x_{n+1}= 1-\alpha x_n^2+y_n
\end{equation}
\begin{equation}
    y_{n+1} = \beta x_n
\end{equation}

The traditional Henon map expresses chaotic behavior with parameter values of $\alpha = 1.8$ and $\beta = 0.3$. However, varying these parameters will result in different chaotic behaviors. Figure 4 shows the bifurcation diagram of the Henon map.

\begin{figure}[h!]
\centering
\includegraphics[width=0.45\textwidth]{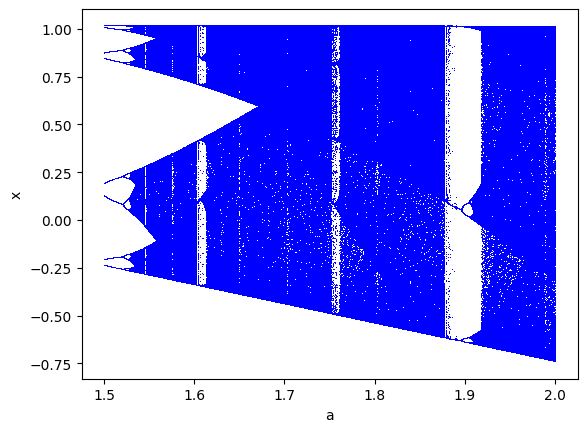}
\caption{ Bifurcation diagram of Henon Map: $\beta$=0.3} 
\label{fig1}
\end{figure}

\subsubsection{Hybrid Chaotic Map}

Hybrid chaotic maps are mathematical constructs that merge various chaotic systems or maps to produce complex and adaptable dynamical patterns \cite{Gao2024CLSM-IEA:Scheme}. By leveraging techniques such as sequential or parallel coupling, these hybrid maps show heightened intricacy and volatility beyond what their individual components provide. The integration of these maps aims to foster more convoluted dynamics, amplify randomness, or enhance specific attributes of chaotic behavior. The logistic-sine map is utilized in our model to generate the sequence used in the confusion region. The system equation of the logistic-sin map is as follows.

\begin{equation}
    x_{n+1}= rx_n(1-x_n)+4r.sin(\frac{\pi x_n}{4})
\end{equation}

When r varies from 0.6 to 1.2, the map shows chaotic behavior. Figure 5 illustrates the Hybrid map's bifurcation diagram.

\begin{figure}[h!]
\centering
\includegraphics[width=0.45\textwidth]{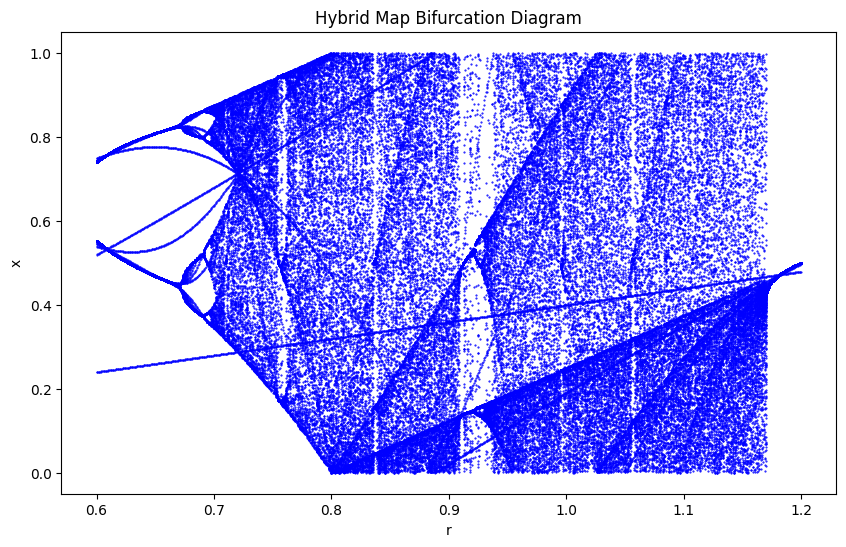}
\caption{Hybrid Logistic-Sin Map Bifurcation diagram} 
\label{fig1}
\end{figure}

\subsubsection{3D Quantum Logistic Chaotic Map}

Similar to traditional 3D chaotic maps \cite{Zhang2022MultipleAlgorithm}, quantum maps also demonstrate heightened chaos. Yet, the quantum variant of the system shows unique quantum characteristics, including interference and entanglement, stemming from its quantum nature. Ongoing researchers aim to delve into its applications and properties, fostering novel insights. An optimistic avenue of investigation involves leveraging the system for quantum chaos-based cryptography, utilizing its chaotic dynamics to generate secure cryptographic keys. 3D equations 4-6 are used to represent the 3D quantum logistic chaotic system.

\begin{equation}
    x_{n+1}=\eta (x_n - |x_n|^2) - \eta y_n
    \end{equation}
\begin{equation}
    y_{n+1}=-y_ne^{-2\gamma}+e^{-\gamma}\eta[(2-x_n-x_n^*)y_n-x_nz_n^*-x_n^*z_n]
    \end{equation}
\begin{equation}
    z_{n+1}=-z_ne^{-2\gamma}+e^{-\gamma}\eta[2(1-x_n^*)z_n-2x_ny_n-x_n]
\end{equation}

In this equation, $\gamma$ denotes the dissipation parameter, $\eta$ stands for the control parameter, and $x_n^*, y_n^*$ represent the complex conjugates of $x_n, y_n$. The state of equations is chaotic when $\eta=4$,$x_n \in (0,1]$, $y_n \in (0,0.1]$, $z_n \in (0,0.2]$, $\gamma \in [6,+\infty]$. Figure 6 shows the phase diagrams of the used quantum logistic map to demonstrate its randomness.

\begin{figure}[h!]
     \centering
     \begin{subfigure}[b]{0.3\textwidth}
         \centering
         \includegraphics[width=1\textwidth]{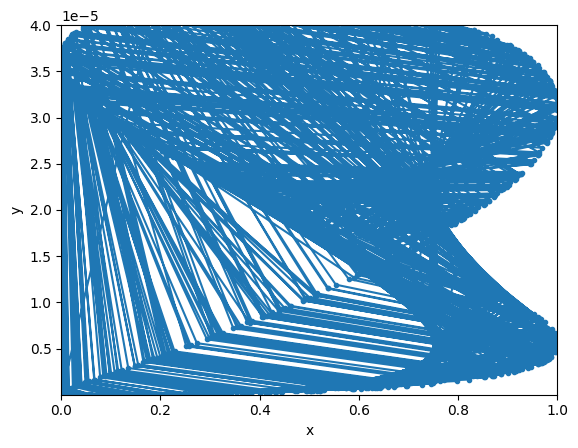}
         \caption{xy plane}
     \end{subfigure}
     \hfill
     \begin{subfigure}[b]{0.3\textwidth}
         \centering
         \includegraphics[width=1\textwidth]{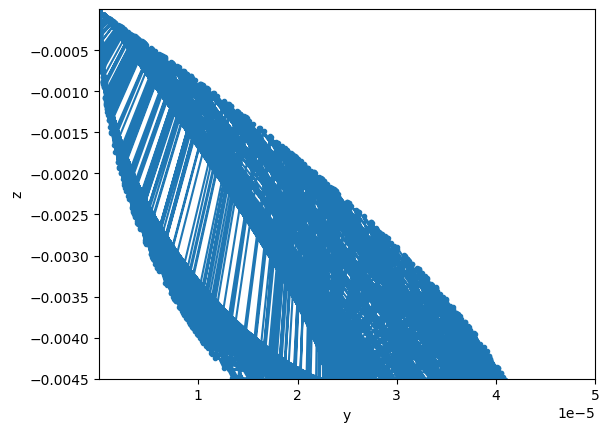}
         \caption{yz plane}
     \end{subfigure}
     \hfill
     \begin{subfigure}[b]{0.3\textwidth}
         \centering
         \includegraphics[width=1\textwidth]{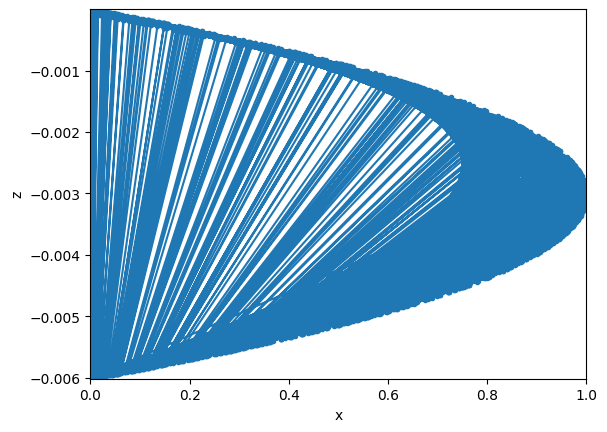}
         \caption{xz plane}
     \end{subfigure}
        \caption{Phase diagrams of Quantum Logistic Map}
\end{figure}

Comparing the 3D quantum logistic map to the 3D Lorenz chaotic map reveals distinct characteristics that influence their suitability for different applications. While the 3D quantum logistic map operates within the framework of quantum mechanics, leveraging properties like superposition and entanglement, the 3D Lorenz chaotic map is rooted in classical chaos theory. This contrast results in differences in complexity, nonlinearity, and computational efficiency between the two models.

\subsection{Quantum Operations}

In the evolving quantum age, leveraging qubits rather than using classical bits in image encryption provides enhanced security through the principles of quantum mechanics. Quantum operators such as the Hadamard and CNOT gates offer unique benefits: the Hadamard gate allows for the formation of superposition states, enabling efficient encoding and manipulation of information, while the CNOT gate facilitates controlled operations between qubits, essential for implementing complex encryption algorithms \cite{Houshmand2024OptimizedImages}\cite{Wang2022AnChaos}. These quantum gates improve the performance of encryption schemes, offering increased randomness and variability in the encryption process, thus strengthening the security posture against cryptographic attacks. Additionally, the inherent properties of qubits, including entanglement and the ability to represent and process information in a fundamentally different way than classical bits, contribute to the robustness of image encryption schemes in the quantum domain. Through the integration of quantum operators, image encryption systems can achieve higher levels of security and resilience against both classical and potential quantum threats.

\subsubsection{Hadamard Gate}
\begin{figure}[h!]
     \centering
     \begin{subfigure}[b]{0.3\textwidth}
         \centering
         \includegraphics[width=1\textwidth]{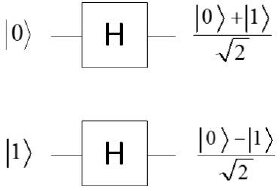}
         \caption{}
     \end{subfigure}
     \hfill
     \begin{subfigure}[b]{0.3\textwidth}
         \centering
         \includegraphics[width=1\textwidth]{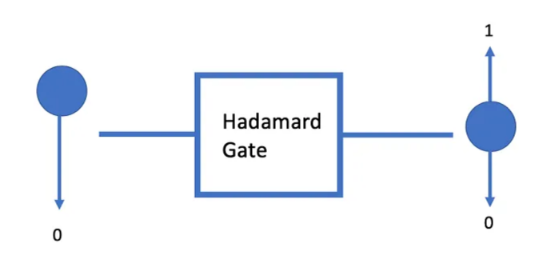}
         \caption{}
     \end{subfigure}
     \hfill
     \begin{subfigure}[b]{0.3\textwidth}
         \centering
         \includegraphics[width=1\textwidth]{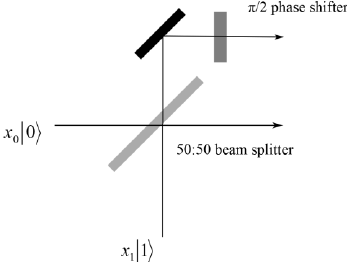}
         \caption{}
     \end{subfigure}
        \caption{Hadamard Gate Representations}
\end{figure}

The Hadamard gate serves as a fundamental quantum gate for generating superposition states. Upon application to a qubit, it transforms a basis state $|0$⟩ into an equal superposition of $|0$⟩ and $|1$⟩, and vice versa. Mathematically, it represents a rotation of the qubit's state vector by 90 degrees around the x-axis of the Bloch sphere. The Hadamard gate plays a crucial role in quantum algorithms, such as quantum Fourier transforms and quantum search algorithms, by generating and manipulating superposition states. Its application enables quantum computers to efficiently process information in parallel and perform certain calculations more effectively than classical computers. Quantum circuit of the $\mathcal{H}$ gate is shown in Figure 7(a), and the Hadamard matrix is described by

\begin{equation}
    \mathcal{H} = \frac{1}{\sqrt{2}}\begin{bmatrix}
1 & 1 \\
1 & -1 
\end{bmatrix}
\end{equation}

Figure 7(b) illustrates that when a qubit is set to the state $|0$〉 at first, subject to the $\mathcal{H}$ gate operation, it moves a superposition state where the probabilities of estimating 0 and 1 are equal. The optical implementation of the $\mathcal{H}$ gate is shown in Figure 7(c).

\subsubsection{CNOT Gate}

The Controlled-NOT (CNOT) gate is a basic gate used for entangling qubits and implementing controlled operations in quantum computing. The CNOT gate functions on two qubits, with one serving as the control and the other as the target. When the control qubit is in state $|1$⟩, the CNOT gate flips the state of the target qubit. However, if the control qubit is in state $|0$⟩, the target qubit remains unaffected. The CNOT gate is essential for creating entanglement between qubits, enabling the execution of quantum algorithms such as quantum teleportation and error correction. Its versatility makes it a cornerstone in quantum computation and communication protocols. It is the same as the XOR operation in classical computation. The CNOT matrix representation is shown in Equation 8. Figure 8 visualizes the quantum circuit of the CNOT gate.

\begin{equation}
    \mathcal{CNOT} = \begin{bmatrix}
1 & 0 & 0 & 0 \\
0 & 1 & 0 & 0 \\
0 & 0 & 0 & 1 \\
0 & 0 & 1 & 0 
\end{bmatrix}
\end{equation}

\begin{figure}[h!]
\centering
\includegraphics[width=0.45\textwidth]{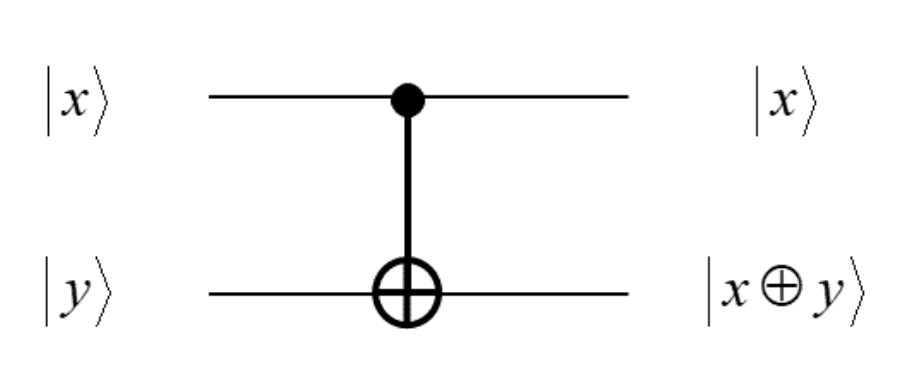}
\caption{CNOT Quantum Circuit} 
\label{fig1}
\end{figure}

\subsection{DNA encoding} 

 \begin{table}[h!]
\caption{DNA Encoding Rules}
\label{tab1e}
\centering
\begin{tabular}{p{1cm} p{1cm} p{1cm} p{1cm} p{1cm} p{1cm} p{1cm} p{1cm} p{1cm}}
\hline
\textbf{Bits} & \textbf{i} & \textbf{ii} & \textbf{iii} & \textbf{iv} & \textbf{v} & \textbf{vi} & \textbf{vii} & \textbf{viii} \\ [0.5ex]
\hline 
00 & A & C & T & A & G & C & T & G\\
01 & G & A & G & C & T & T & G & A\\
10 & C & T & C & G & A & A & C & T\\
11 & T & G & A & T & C & G & A & C\\
\hline
\end{tabular}
\end{table}

DNA encoding involves representing digital data using the four nucleotide bases of DNA which are adenine(A), guanine(G), cytosine(C), and thymine(T), instead of binary digits. This technique utilizes the unique properties of DNA molecules, such as their high storage capacity and stability, for data storage and computation \cite{Li2023ADiffusion}. A DNA sequence could be generated from binary data, and Table 2 lists eight DNA encoding techniques. DNA computing holds significant promise in image cryptography due to its ability to leverage massive parallelism, high data density, robustness, and unique security properties. By encoding image data into DNA sequences, DNA-based encryption schemes can efficiently process large-scale image data while ensuring secure storage and transmission. The inherent robustness of DNA sequences enhances the resilience of image cryptography systems against data corruption and tampering, while their bio-inspired nature offers novel approaches to encryption. Arithmetic and logical operations can also be executed on DNA-encoded information, with the truth table of the DNA XOR operation presented in Table 3.\\

\begin{table}[h!]
\caption{DNA XOR truth\_table}
\label{tab1e}
\centering
\begin{tabular}{p{1cm} p{1cm} p{1cm} p{1cm} p{1cm}}
\hline
\textbf{XOR} & \textbf{A} & \textbf{G} & \textbf{T} & \textbf{C} \\ [0.5ex]
\hline 
\textbf{T} & T & C & A & G \\
\textbf{G} & G & A & C & T \\
\textbf{A} & A & G & T & C \\
\textbf{C} & C & T & G & A \\
\hline
\end{tabular}
\end{table}

\section{Proposed Encryption Model}
The proposed QMedShield: a Quantum chaos-based image encryption model, is explained in detail in this module. The detailed introduced encryption flow is shown in Figure 9. Key Management Centre (KMC) generates the keys to share with users.\\

\begin{figure}[h!]
\centering
\includegraphics[width=1.0\textwidth]{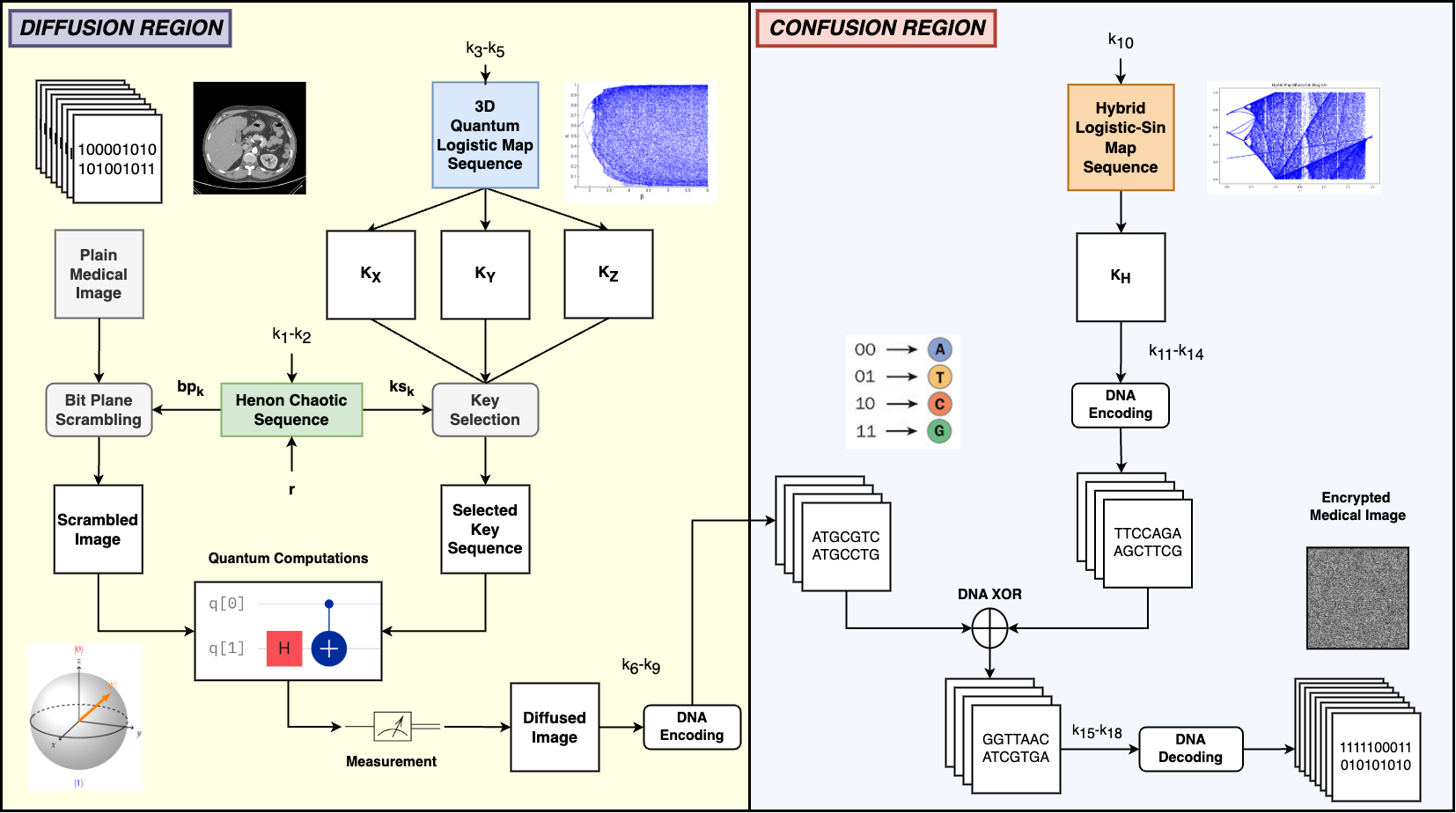}
\caption{Flow Diagram of the Image Encryption} 
\label{fig1}
\end{figure}

The proposed model is founded on the permutation-diffusion model of Shannon, which forms the fundamental basis. QMedShield has two regions to be processed. One is the diffusion region where image bit planes are scrambled and pixel values are diffused using a generated key with quantum operations. Another one is the confusion region, where pixel substitution using DNA encoding and XOR operations takes place. The detailed steps are explained as follows.  

\subsection{Algorithm of QMedShield}
Secret key ($K = \{k_1, k_2, k_3, k_4,....,k_{18}\}$) Medical Image ($Med_i$) from KMC are given to the authorized image owners who can encrypt the medical image data. The resultant Cipher Medical Image $CMed_i$ will be offloaded to the cloud for secure storage.\\

\textbf{}\\
\textbf{Step 1:} Grayscale input medical image is read as in 8-bit plane format, $Med_i =\{b_{i0},b_{i1},b_{i2},...,b_{i7}\}$.\\
\textbf{Step 2:} Then, the bit plane scrambling key ($bp_{k}$) and key selection key ($ks_{k}$) are derived by giving initial parameters $x_0 = k_1, y_0 = k_2$ and seed $r$ to the Henon map as described in Equations 1-2. It is assumed that the value of $r$ is confidentially shared by the receiver and sender of the medical image .\\
\textbf{Step 3:} The major 3D quantum logistic map sequences are generated from the chaotic system represented by equations 4-6 by feeding the initial parameters $x_0 = k_3, y_0 = k_4, z_0 = k_5$. Each sequence matches the size of an image $Med_i$. These matrices are the diffusion keys and they are computed leveraging the given equations 9-11. where $(\epsilon_1, \epsilon_2)$ denote two large prime numbers and $x_i, y_i, z_i$ represents random sequences, generated using 3D quantum logistic map.\\

\begin{equation}
    K_X = mod( floor(\epsilon_1  x_i + \epsilon_2), 256)
\end{equation}
\begin{equation}
    K_Y = mod( floor(\epsilon_1  y_i + \epsilon_2), 256)
\end{equation}
\begin{equation}
    K_Z = mod( floor(\epsilon_1 z_i + \epsilon_2), 256)
\end{equation}
\textbf{Step 4:} $Med_i$ is being scrambled  using $bp_{k}$, outputs scrambled image $SI_i = \{SI_{i0},SI_{i1}, SI_{i2}, SI_{i3},....., SI_{i7}\}$. Meanwhile, from the three quantum logistic keys, one key $K_S$ is randomly selected based on $ks_{k}$.\\
\textbf{Step 5:} In a quantum circuit, the chosen key $K_S$ and the scrambled image $SI_i$ undergo $\mathcal{H}$ transformation with CNOT gate to create an entangled qubit. This process comprises applying a $\mathcal{H}$ gate and then a CNOT gate to convoy all values in the resulting matrices to a superposition state and establish entanglement. To effectively achieve pixel-level diffusion, an XOR function is applied with the corresponding qubits. Corresponding quantum circuit diagrams can be found in sections 3.3.1 and 3.3.2.\\
\textbf{Step 6:} The quantum state of the diffused grayscale pixel of the image is collapsed into their classical versions by the quantum measurement process, denoted as $C_i$.\\
\textbf{Step 7:} The obtained diffused image $C_i$ is transformed into DNA encoded 4 planes $DN_i = \{DN_{i0}, DN_{i1}, DN_{i2}, DN_{i3}\}$ using $k_6-k_9$.\\
\textbf{Step 8:} The hybrid logistic-sine map sequence $K_H$ is generated using equation 5 by giving the initial parameter $x_0=k_{10}$. The sequence length is the same as the size of $Med_i$.\\
\textbf{Step 9:} The generated key $K_H$ is converted into DNA encoded 4 planes $DK_i = \{DK_{i0}, DK_{i1}, DK_{i2}, DK_{i3}\}$ using $k_{11}-k_{14}$.\\
\textbf{Step 10:} $DX_i = DNA\_XOR(DN_{ik},DK_{ik})$, $k = 0,1,2,3$.\\
\textbf{Step 11:} The DNA XOR image $DX_i$ is DNA decoded using $k_{15}-k_{18}$ and produces the encrypted medical image $CMed_i$.\\

In the QMedShield, harnessing qubits instead of classical bits in image encryption enhances security through quantum operators like the Hadamard and CNOT gates. The Hadamard gate enables superposition states, facilitating efficient encoding, while the CNOT gate allows controlled operations between qubits, crucial for complex encryption algorithms. These quantum gates reinforce encryption schemes by introducing randomness and variability, thereby fortifying security against attacks. Moreover, the unique properties of qubits, including entanglement, provide resilience against classical and potential quantum threats, enhancing the security of image encryption within the quantum domain. The decryption process involves reversing the given steps. The original medical image $Med$ can be reconstructed using the equation below and the process flow is illustrated in Figure 10. 
\begin{equation}
    Med_i = Decryption(CMed_i,K,r)
\end{equation}

\begin{figure}[h!]
\centering
\includegraphics[width=1.0\textwidth]{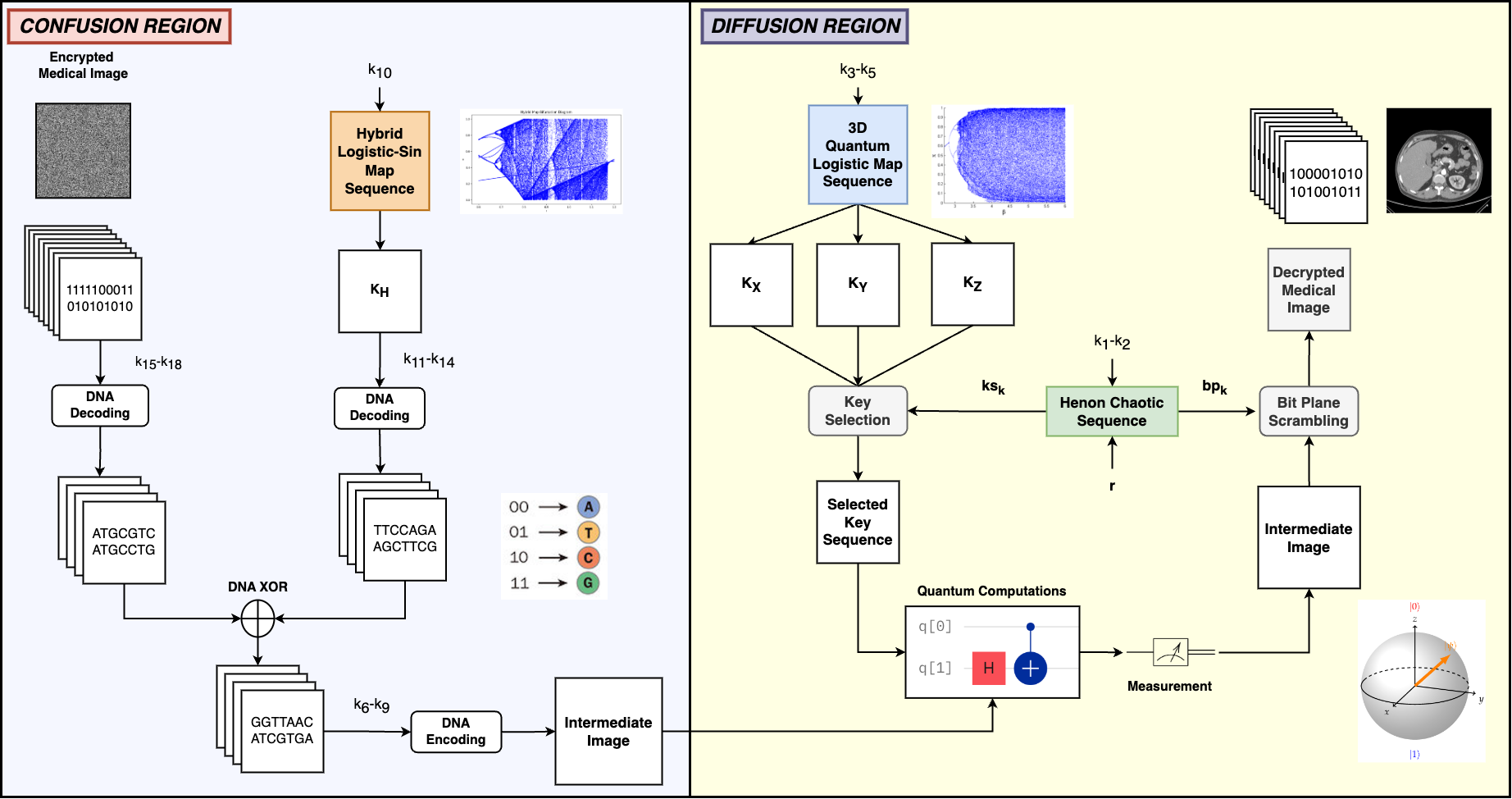}
\caption{Flow Diagram of the Image Encryption} 
\label{fig1}
\end{figure}

\section{Experimental Results and Analysis}

This section is dedicated to analyzing the proposed encryption model. We have outlined the experimental setup, including the medical dataset used for analysis, and presented various experimental results accordingly. This includes analyses such as statistical attack analysis, key security analysis, analyses of resistance to chosen plaintext and known plaintext attacks, and differential attack analysis.

\subsection{Experimental Setup and Dataset}

The designed encryption model was implemented on a PC with an Intel Xeon processor, 64 GB RAM, 16 GB of memory with an NVIDIA Quadro P5000 GPU, and a 64-bit Windows operating system. Python OpenCV libraries and Qiskit have been used in the development of the entire system. To experiment and evaluate the proposed algorithm, three different medical image datasets are chosen. Details of the dataset have been explained briefly here,

\begin{itemize}
    \item \textbf{Brain Tumor MRI Dataset (BMRI) \cite{MsoudNickparvar2021BrainDataset}:} The three datasets below are combined to create this dataset: figshare, SARTAJ, Br35H. There are 7023 MRI images of the human brain in this collection, divided into 4 categories: pituitary, glioma, meningioma, and no tumor. Images categorized as the 'no tumor' class were obtained from the Br35H dataset.
    \item \textbf{Chest X-ray Dataset (CXR) \cite{WangChestX-ray8:Diseases}:} This dataset originates from the NIH,  which is the largest chest radiograph data set. From 30,805 special patients, 112,120 frontal X-ray images are collected. Each X-ray is linked to the associated text disease label, which is drawn from the relevant radiological reports using an NLP algorithm. 
    \item \textbf{Lung Cancer CT Dataset (LCT) \cite{Nitha2023ExtRanFS:Selector}:} From different specialist hospitals, the IQ-OTH/NCCD lung cancer dataset was collected over three months in the fall of 2019. It comprises CT scans from patients with lung cancer in different stages and healthy subjects, totaling 1190 images from 110 cases. The dataset, marked by oncologists and radiologists, categorizes cases into three classes: normal (55 cases), benign (15 cases), and malignant (40 cases).
    
\end{itemize}

The proposed QMedShield's security is evaluated using a number of metrics and proved that it is resistant to various cryptographic attacks including brute-force attacks, statistical attacks, histogram attacks, and differential attacks. Throughout the section, 6 sample medical images $BMRI_1, BMRI_2, CXR_1, CXR_2, LCT_1, LCT_2$ are taken (2 images from each dataset) to show the performance comparison. The selection of MRI, X-ray, and CT images for the encryption task aims to demonstrate the versatility and effectiveness of our model across various imaging modalities, showcasing its applicability and robustness in diverse clinical scenarios. Figure 11 shows the selected sample medical images and their corresponding encrypted images.

\begin{figure}[h!]
     \centering
     \begin{subfigure}[b]{0.3\textwidth}
         \centering
         \includegraphics[width=0.65\textwidth]{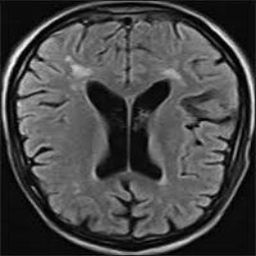}
         \caption{$BMRI_1$}
     \end{subfigure}
     \hfill
     \begin{subfigure}[b]{0.3\textwidth}
         \centering
         \includegraphics[width=0.65\textwidth]{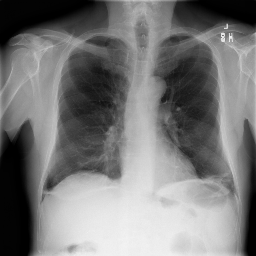}
         \caption{$CXR_1$}
     \end{subfigure}
     \hfill
     \begin{subfigure}[b]{0.3\textwidth}
         \centering
         \includegraphics[width=0.65\textwidth]{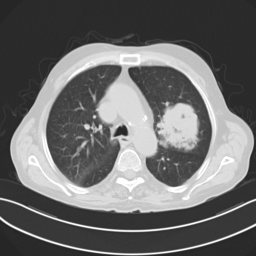}
         \caption{$LCT_1$}
     \end{subfigure}
     \begin{subfigure}[b]{0.3\textwidth}
         \centering
         \includegraphics[width=0.65\textwidth]{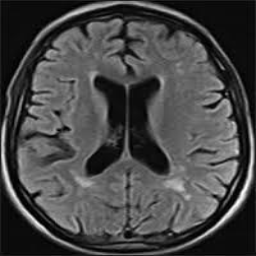}
         \caption{$BMRI_2$}
     \end{subfigure}
     \hfill
     \begin{subfigure}[b]{0.3\textwidth}
         \centering
         \includegraphics[width=0.65\textwidth]{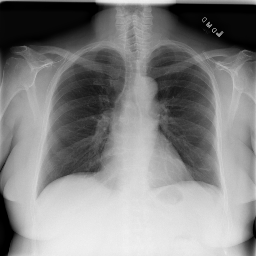}
         \caption{$CXR_2$}
     \end{subfigure}
     \hfill
     \begin{subfigure}[b]{0.3\textwidth}
         \centering
         \includegraphics[width=0.65\textwidth]{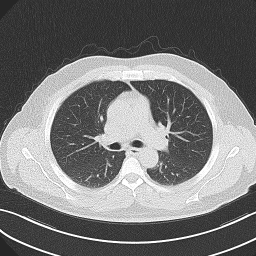}
         \caption{$LCT_2$}
     \end{subfigure}
     \begin{subfigure}[b]{0.3\textwidth}
         \centering
         \includegraphics[width=0.65\textwidth]{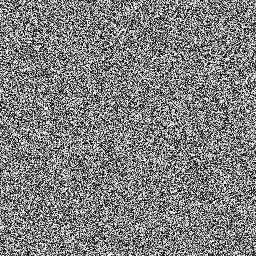}
         \caption{$E(BMRI_1)$}
     \end{subfigure}
     \hfill
     \begin{subfigure}[b]{0.3\textwidth}
         \centering
         \includegraphics[width=0.65\textwidth]{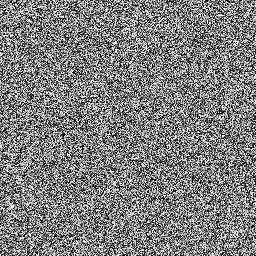}
         \caption{$E(CXR_1)$}
     \end{subfigure}
     \hfill
     \begin{subfigure}[b]{0.3\textwidth}
         \centering
         \includegraphics[width=0.65\textwidth]{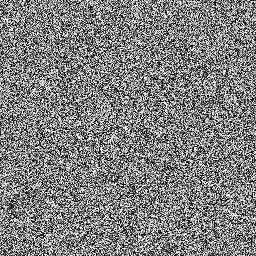}
         \caption{$E(LCT_1)$}
     \end{subfigure}
     \begin{subfigure}[b]{0.3\textwidth}
         \centering
         \includegraphics[width=0.65\textwidth]{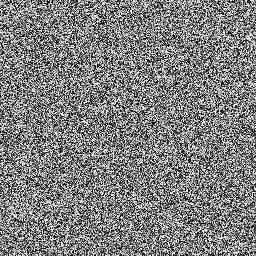}
         \caption{$E(BMRI_2)$}
     \end{subfigure}
     \hfill
     \begin{subfigure}[b]{0.3\textwidth}
         \centering
         \includegraphics[width=0.65\textwidth]{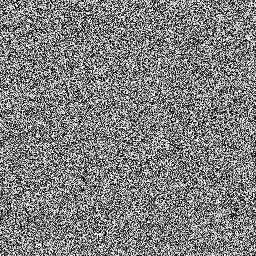}
         \caption{$E(CXR_2)$}
     \end{subfigure}
     \hfill
     \begin{subfigure}[b]{0.3\textwidth}
         \centering
         \includegraphics[width=0.65\textwidth]{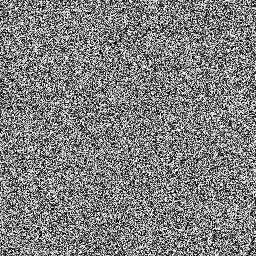}
         \caption{$E(LCT_2)$}
     \end{subfigure}
        \caption{Selected sample plain medical images and the corresponding encrypted cipher images}
\end{figure}

\subsection{Key Space Analysis}

In contemporary cryptography, the magnitude of the key space plays an important role in discovering the resilience of a crypto model against brute-force attacks \cite{Wan2020ACoding}. To thwart such attacks, robust passphrases or passwords and encryption models boasting adequately expansive key capacities are indispensable. It is commonly recommended to maintain a space of at least $2^{128}$ to withstand brute-force endeavors \cite{AmaithiRajan2023SecureEncoding}. Typically, a larger key space renders it increasingly arduous to deduce the exact security key through brute force, thereby augmenting the security of the model. The key volume computation hinges on the count of unique keys employed in both the confusion and diffusion stages. In a chaotic map, initial values and control parameters serve as keys. For the proposed QMedshield scheme, the private key encompasses a collection of initial conditions and control parameters, delineated as follows: 1) Henon chaotic map has ($x_0,y_0,\alpha, \beta$). 2) Hybrid logistic-sin map has ($x_0,r$). 3) 3D quantum logistic map have ($x_0,y_0,z_0, \eta, \gamma$). The proposed encryption scheme has 11 keys used to generate chaotic sequences that are more sensitive and have large key space. Let's say, the precision is $10^{-14}$ (i.e., double precision ($2^{52}$)), then the size of key space will be equal to \begin{math}2^{52}*11 = 2^{572} > 2^{128}\end{math}. 
The findings indicate that the method is significantly resilient to brute-force attacks and demonstrate the difficulty of successfully breaching it through this method.
\begin{figure}[h!]
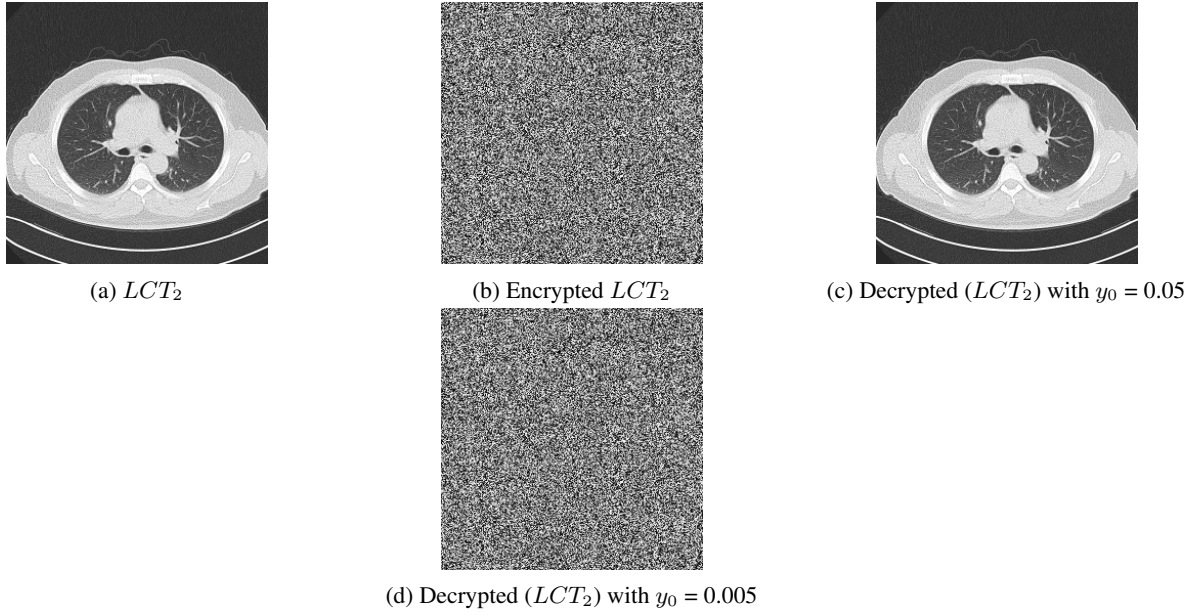

     \centering
     \begin{subfigure}[b]{0.3\textwidth}
         \centering
         \includegraphics[width=0.7\textwidth]{LCT2.png}
         \caption{$LCT_2$}
     \end{subfigure}
     \hfill
     \begin{subfigure}[b]{0.3\textwidth}
         \centering
         \includegraphics[width=0.7\textwidth]{LCT2_Enc.png}
         \caption{Encrypted $LCT_2$}
     \end{subfigure}
     \hfill
     \begin{subfigure}[b]{0.3\textwidth}
         \centering
         \includegraphics[width=0.7\textwidth]{LCT2_Dec.png}
         \caption{Decrypted ($LCT_2$) with $y_0$ = 0.05}
     \end{subfigure}
      \begin{subfigure}[b]{0.3\textwidth}
         \centering
         \includegraphics[width=0.7\textwidth]{LCT2_Wrong.png}
         \caption{Decrypted ($LCT_2$) with $y_0$ = 0.005}
     \end{subfigure}
        \caption{Key Sensitivity Analysis}
\end{figure}

\subsection{Key Sensitivity Analysis}

The encryption algorithm ought to be sensitive to the key. Even minor alterations to the key should result in an entirely different cipher image. The proposed QMedShield demonstrates a high level of sensitivity to the key. The original medical image $LCT_2$ is shown in 12(a). $LCT_2$ is encrypted with initialization values for quantum logistic chaotic map $x_0$ = 0.5, $y_0$ = 0.05, and $z_0$ = 0.02. Encrypted $LCT_2$ is shown in 12(b). If we use the same values while decrypting, we will get the exact original medical image $LCT_2$ (Refer 12(c)). Figure 12(d) shows the result of decryption if $y_0$ is given as 0.005. It explains that even a small modification can have a significant impact. Therefore, it ensures guessing the encryption key completely makes it difficult to decrypt.

\subsection{Histogram Analysis}

The statistical properties of the image are represented through the histogram. It counts the number of pixels and primarily displays the distribution of pixel values within the image. A uniform distribution results in a flat histogram, indicating that pixel values are nearly equal throughout the image. Histogram analysis plays a critical role in both image processing and encryption, serving as a key statistical measure to validate encryption scheme security against statistical attacks. Its widespread use underscores its robust defense against such attacks. The histograms of sample plain images are shown in Figure 13(a-f). The corresponding image's encrypted image histograms are displayed in Figure 13(g-l). The pixel values in the cipher image are uniformly distributed. This experiment demonstrates how pixel value distribution in the original image can be effectively hidden by the cipher image. It protects the data from histogram-based statistical attacks.
\begin{figure}
     \centering
     \begin{subfigure}[b]{0.3\textwidth}
         \centering
         \includegraphics[width=1.2\textwidth]{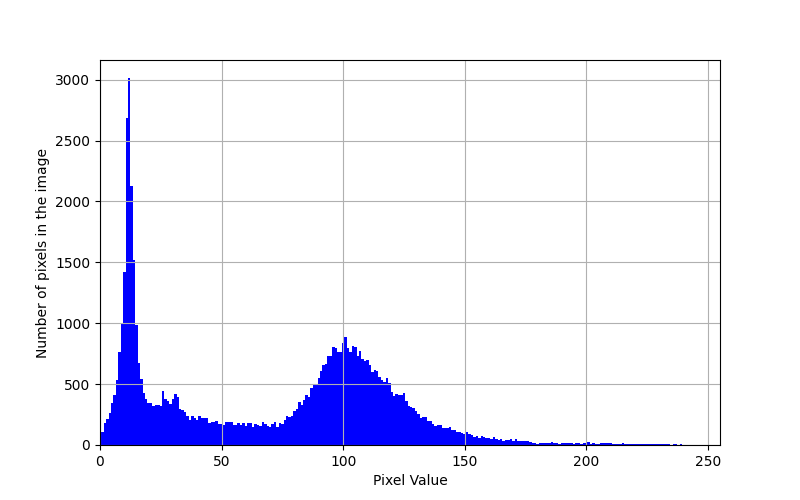}
         \caption{Original($BMRI_1$)}
     \end{subfigure}
     \hfill
     \begin{subfigure}[b]{0.3\textwidth}
         \centering
         \includegraphics[width=1.2\textwidth]{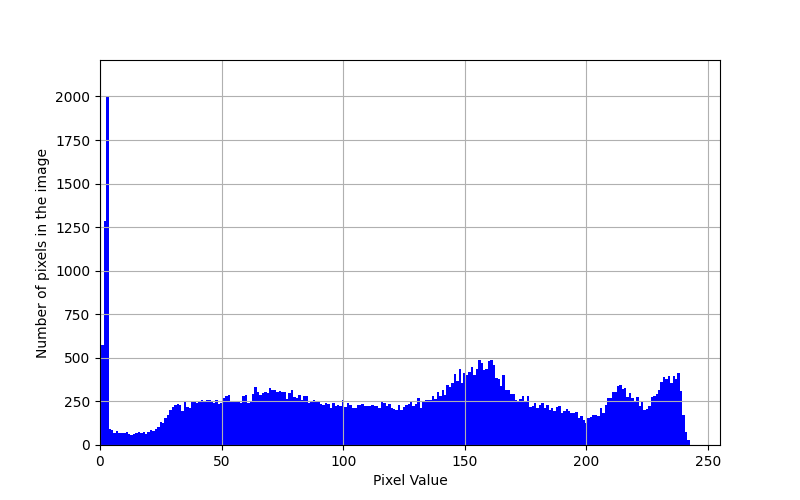}
         \caption{Original($CXR_1$)}
     \end{subfigure}
     \hfill
     \begin{subfigure}[b]{0.3\textwidth}
         \centering
         \includegraphics[width=1.2\textwidth]{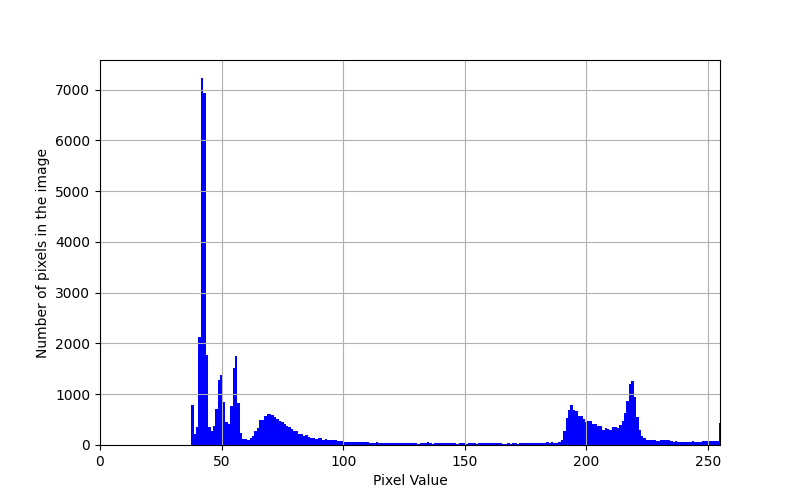}
         \caption{Original($LCT_1$)}
     \end{subfigure}
     \begin{subfigure}[b]{0.3\textwidth}
         \centering
         \includegraphics[width=1.2\textwidth]{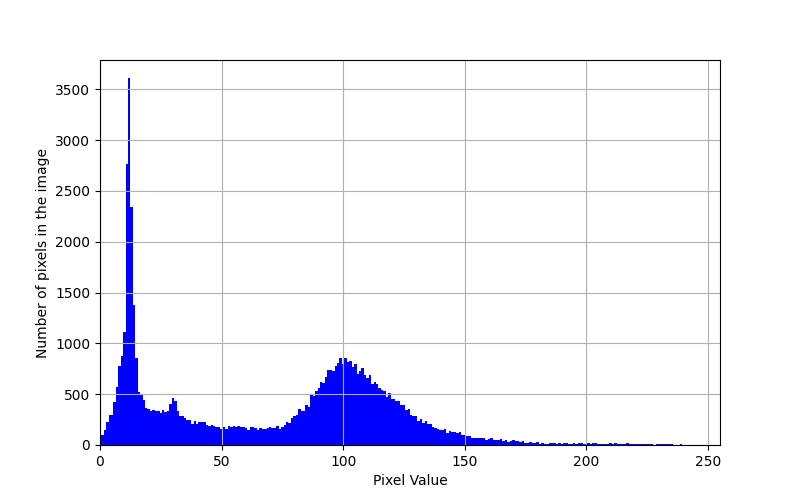}
         \caption{Original($BMRI_2$)}
     \end{subfigure}
     \hfill
     \begin{subfigure}[b]{0.3\textwidth}
         \centering
         \includegraphics[width=1.2\textwidth]{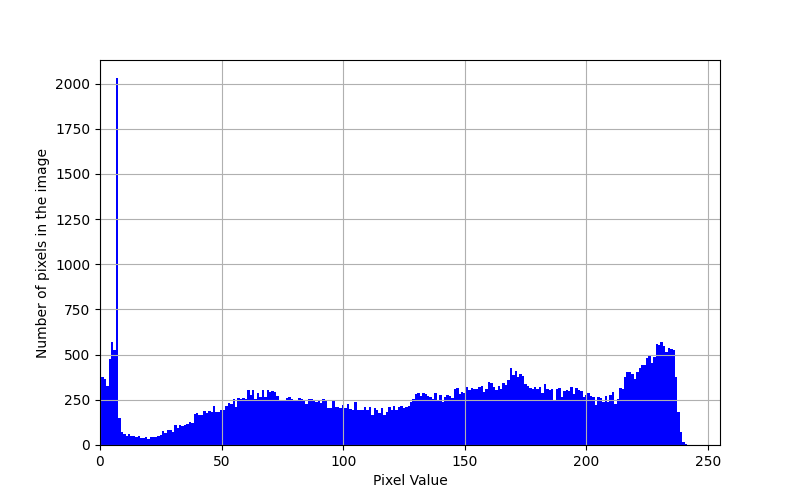}
         \caption{Original($CXR_2$)}
     \end{subfigure}
     \hfill
     \begin{subfigure}[b]{0.3\textwidth}
         \centering
         \includegraphics[width=1.2\textwidth]{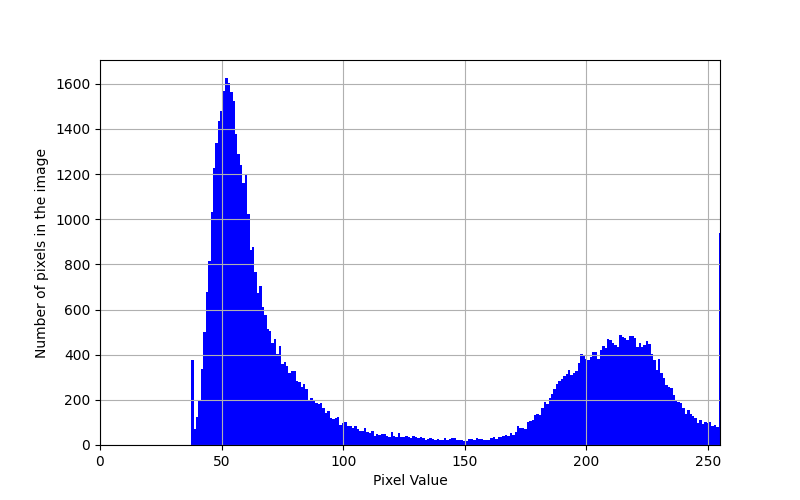}
         \caption{Original($LCT_2$)}
     \end{subfigure}
     \begin{subfigure}[b]{0.3\textwidth}
         \centering
         \includegraphics[width=1.2\textwidth]{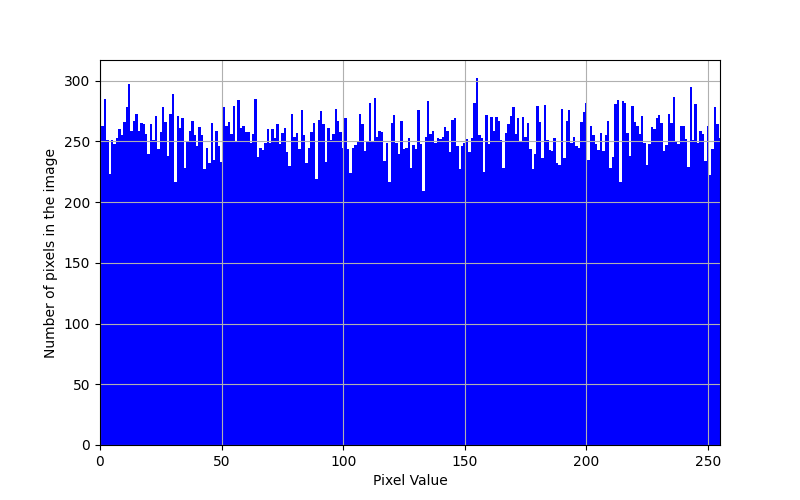}
         \caption{Encrypted($BMRI_1$)}
     \end{subfigure}
     \hfill
     \begin{subfigure}[b]{0.3\textwidth}
         \centering
         \includegraphics[width=1.2\textwidth]{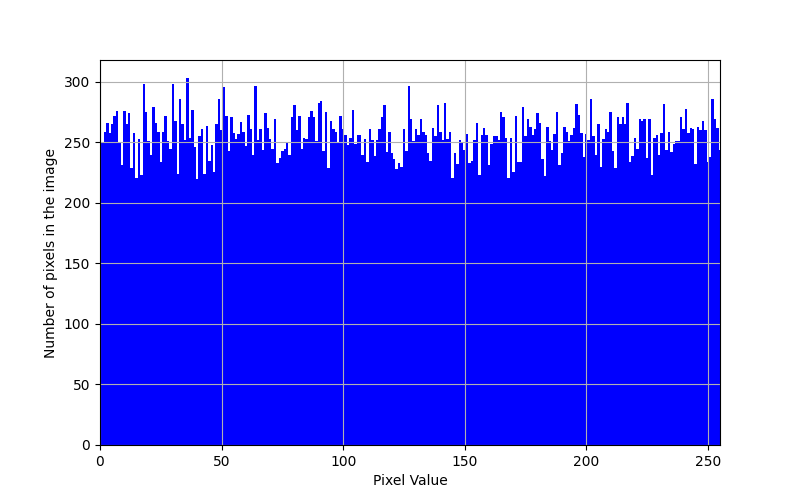}
         \caption{Encrypted($CXR_1$)}
     \end{subfigure}
     \hfill
     \begin{subfigure}[b]{0.3\textwidth}
         \centering
         \includegraphics[width=1.2\textwidth]{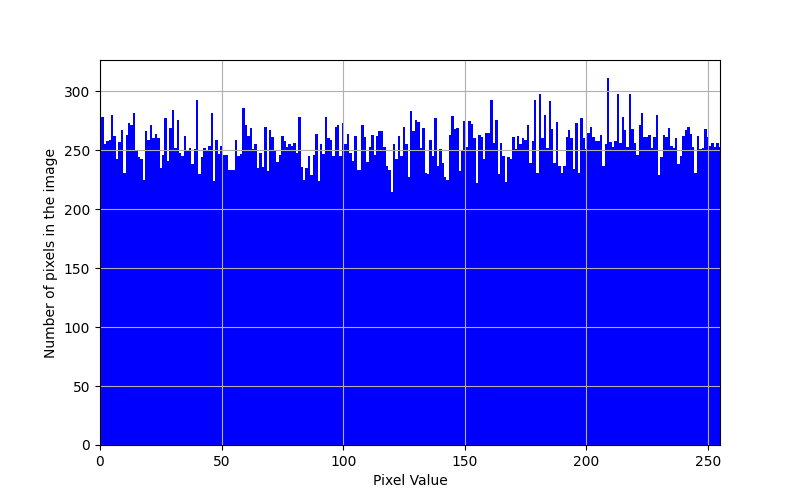}
         \caption{Encrypted($LCT_1$)}
     \end{subfigure}
     \begin{subfigure}[b]{0.3\textwidth}
         \centering
         \includegraphics[width=1.2\textwidth]{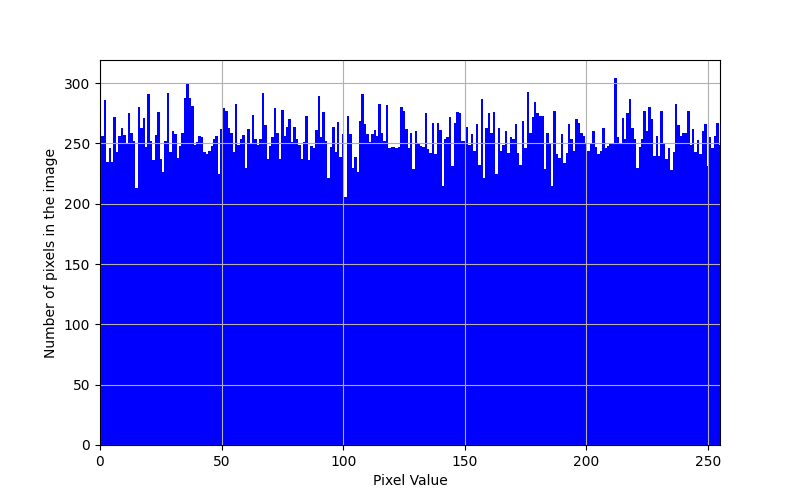}
         \caption{Encrypted($BMRI_2$)}
     \end{subfigure}
     \hfill
     \begin{subfigure}[b]{0.3\textwidth}
         \centering
         \includegraphics[width=1.2\textwidth]{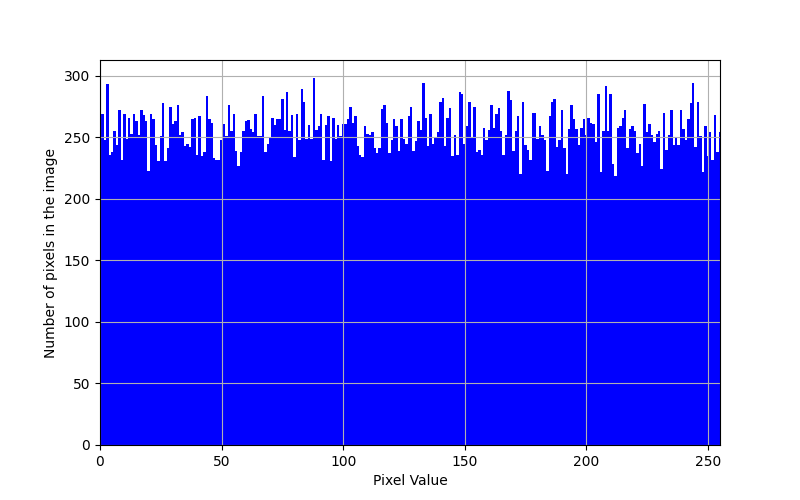}
         \caption{Encrypted($CXR_2$)}
     \end{subfigure}
     \hfill
     \begin{subfigure}[b]{0.3\textwidth}
         \centering
         \includegraphics[width=1.2\textwidth]{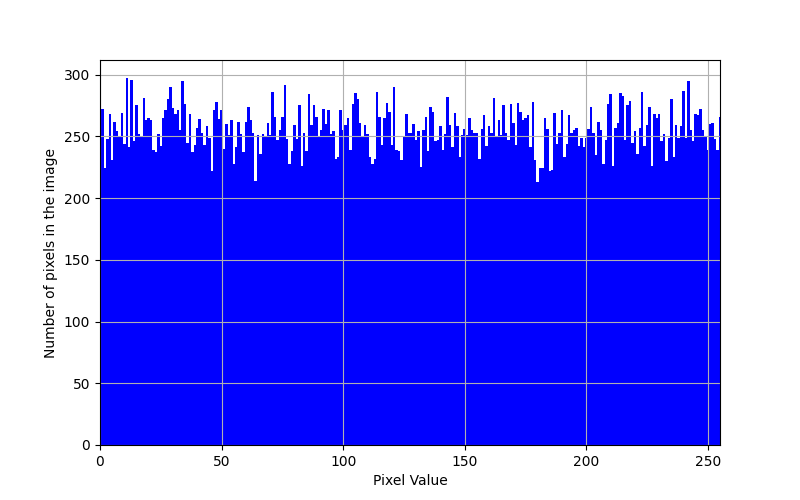}
         \caption{Encrypted($LCT_2$)}
     \end{subfigure}
        \caption{Histogram Analysis}
\end{figure}
\subsection{Chi-Square Test}
The chi-square \begin{math}(\chi^2)\end{math} test is used to assess the histogram's evenness. It is calculated using the following Equation 13. 
\begin{equation}
    \chi^2 = \Sigma_{k=0}^{255} \frac{(OB_k - EX_k)^2}{EX_k}
\end{equation}
In this context, the null hypothesis posits that "Pixels are evenly distributed". A critical value denoted as \begin{math}\chi^2(255,0.05) = 293\end{math}. If the \begin{math}\chi^2\end{math} value is lower than 293, it is concluded that the null hypothesis is valid and accepted.  In Equation 13, $OB, EX$ refers to the observed and expected respectively. Table 4 shows the \begin{math}\chi^2\end{math} test done over the sample images. 
 
\begin{table}[h!]
\caption{\begin{math}\chi^2\end{math} Test}
\label{tab1e}
\centering
\begin{tabular}{cccc}
\hline
\textbf{Image} & \textbf{\begin{math}\chi^2\end{math} Value} & \textbf{Critical Value} & \textbf{Decision (H=0)}  \\ [0.5ex]
\hline 
$BMRI_1$ & 261.63 & 293 & Pass \\
$BMRI_2$ & 273.04 & 293 & Pass \\
$CXR_1$ & 274.57 & 293 & Pass \\
$CXR_2$ & 264.06 & 293 & Pass \\
$LCT_1$ & 265.49 & 293 & Pass \\
$LCT_2$ & 275.76 & 293 & Pass \\
\hline
\end{tabular}
\end{table}

\subsection{Pixel Correlation Analysis}

In image encryption, horizontal, vertical, and diagonal correlation analyses examine the statistical relationships between pixels along different directions within the encrypted image. These analyses provide insights into the spatial distribution of pixel intensities, aiding in the evaluation of encryption effectiveness and the detection of artifacts. By assessing correlation patterns in these directions, encryption practitioners can identify vulnerabilities and optimize algorithms to enhance security and preserve image quality \cite{Zhang2021Multi-imageCoding}.  Equations 14–17 can be used to get the correlation coefficient of an image in any direction. In our case, 3 directions: horizontal, vertical, diagonal. 
\begin{equation}
    Exp(x) = \frac{1}{m}\Sigma_{i=1}^m x_i
\end{equation}
\begin{equation}
    D(x) = \frac{1}{m}\Sigma_{i=1}^m (x_i - Exp(x))^2
\end{equation}
\begin{equation}
    Cov(x,y) = \frac{1}{m}\Sigma_{i=1}^m (x_i - Exp(x))(y_i - Exp(y))
\end{equation}
\begin{equation}
    \rho_{xy} = \frac{Cov(x,y)}{D(x)D(y)}
\end{equation}
The sub-figures 14(a-f) clearly show the randomness in Horizontal, Vertical, and Diagonal directions of the plain and encrypted images of selected medical image samples.
\begin{figure}[h!]
     \centering
     \begin{subfigure}[b]{0.4\textwidth}
         \centering
         \includegraphics[width=1.3\textwidth]{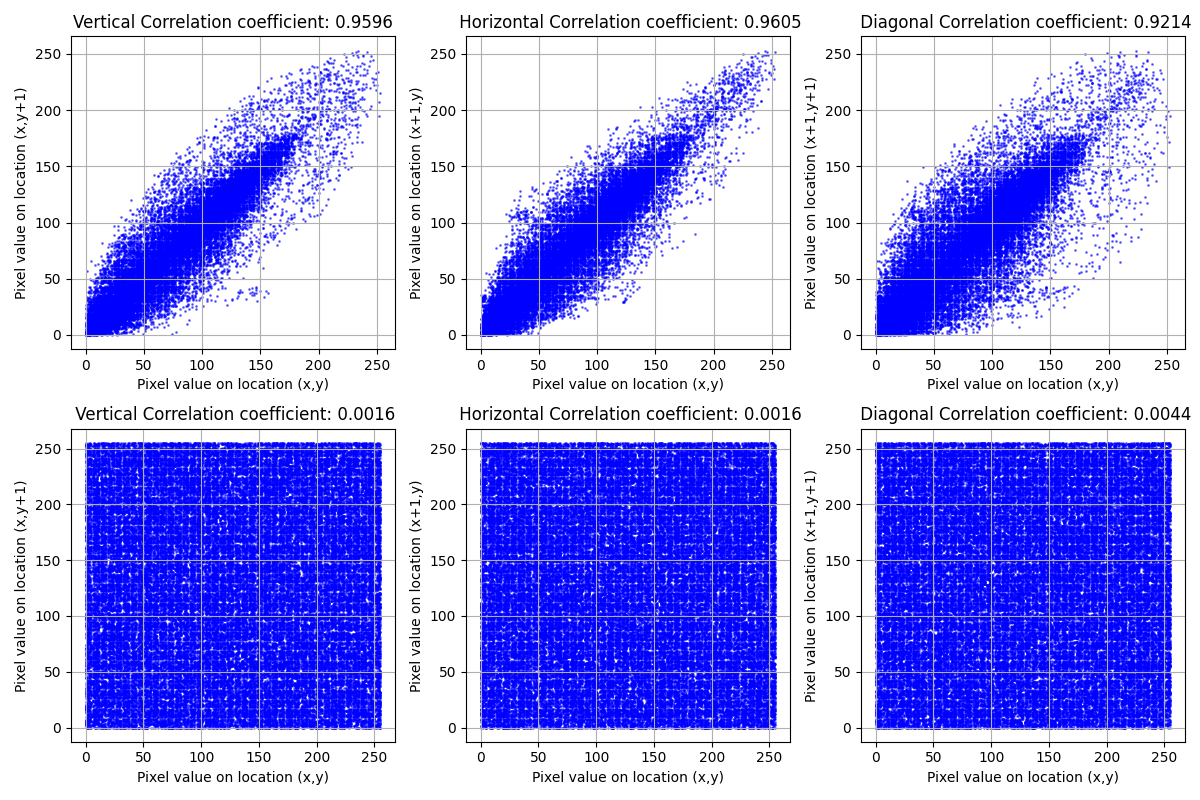}
         \caption{$BMRI_1$}
     \end{subfigure}
     \hfill
      \begin{subfigure}[b]{0.4\textwidth}
         \centering
         \includegraphics[width=1.3\textwidth]{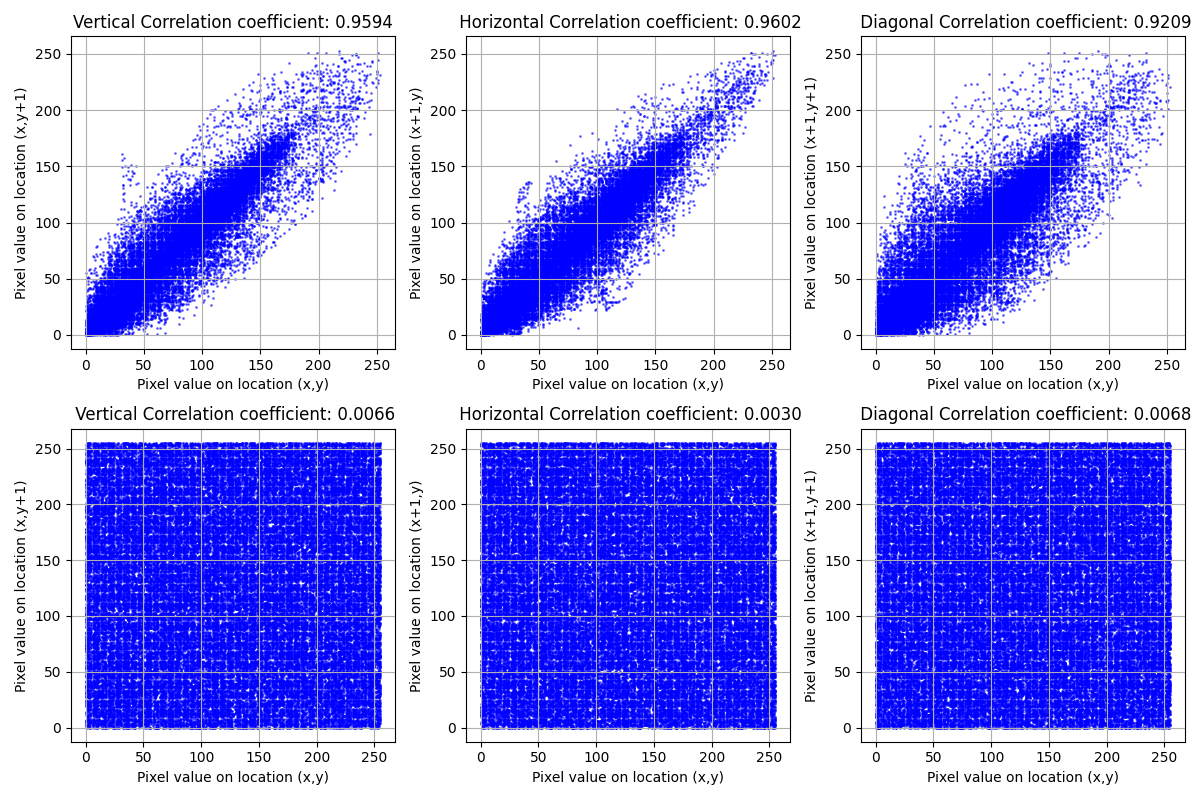}
         \caption{$BMRI_2$}
     \end{subfigure}
     
     \begin{subfigure}[b]{0.4\textwidth}
         \centering
         \includegraphics[width=1.3\textwidth]{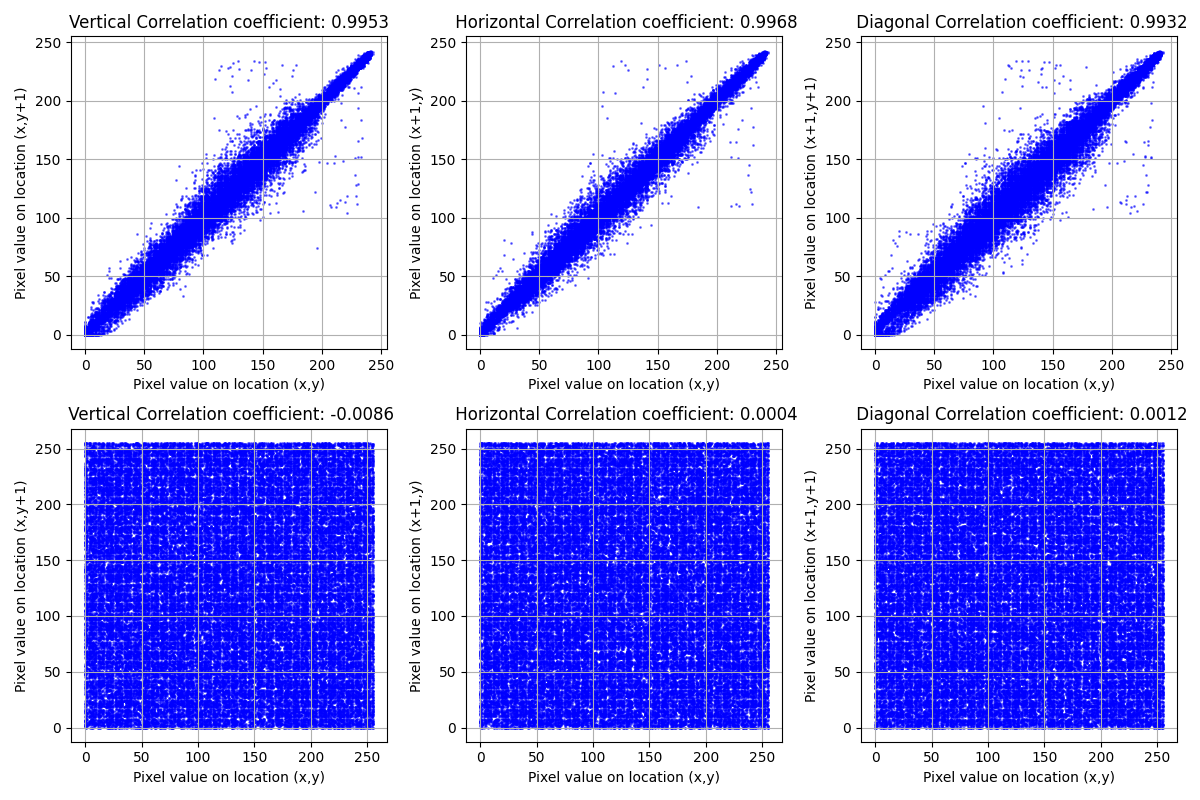}
         \caption{$CXR_1$}
     \end{subfigure}
     \hfill
      \begin{subfigure}[b]{0.4\textwidth}
         \centering
         \includegraphics[width=1.3\textwidth]{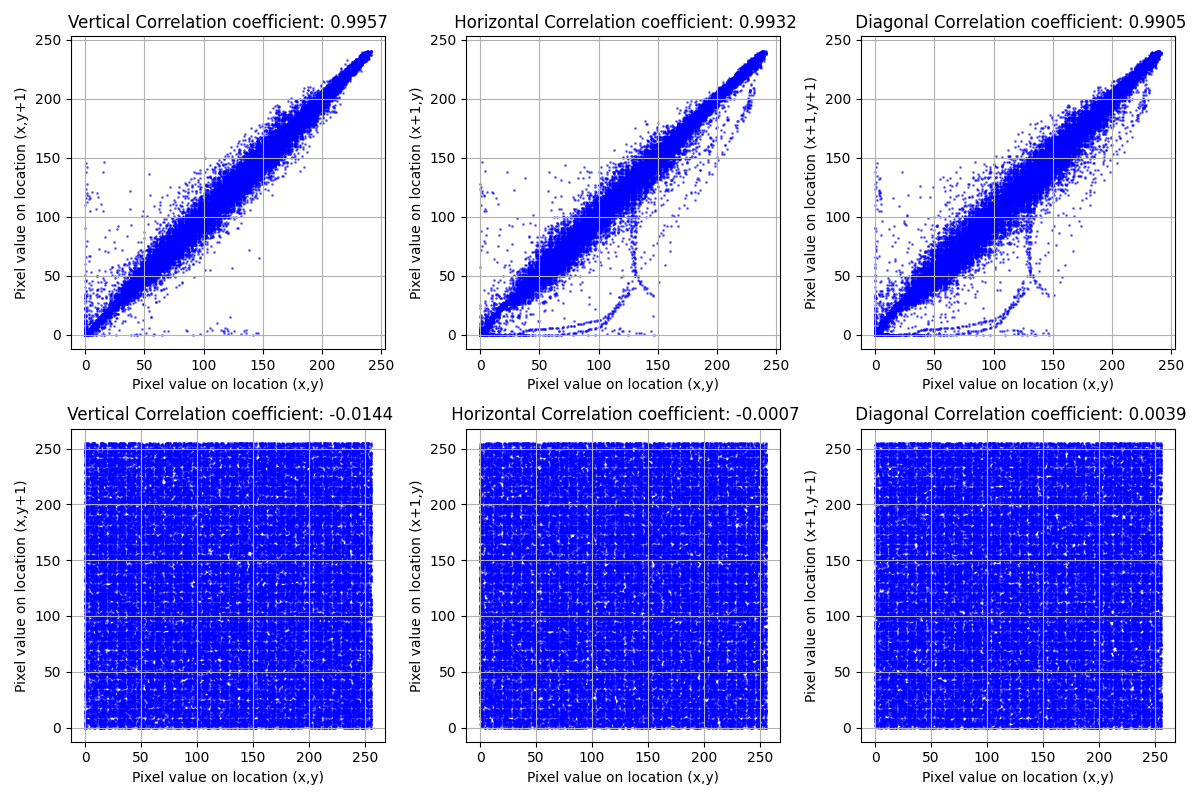}
         \caption{$CXR_2$}
     \end{subfigure}
    
     \begin{subfigure}[b]{0.4\textwidth}
         \centering
         \includegraphics[width=1.3\textwidth]{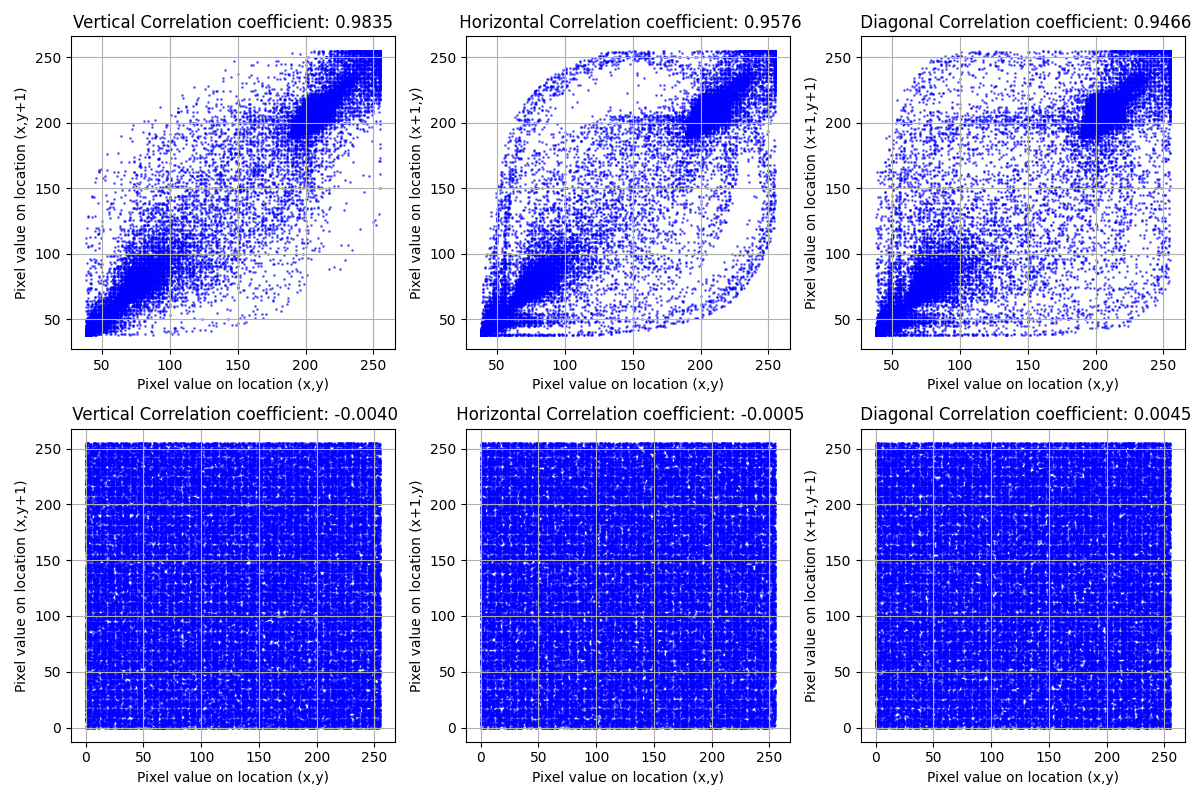}
         \caption{$LCT_1$}
     \end{subfigure}
     \hfill
      \begin{subfigure}[b]{0.4\textwidth}
         \centering
         \includegraphics[width=1.3\textwidth]{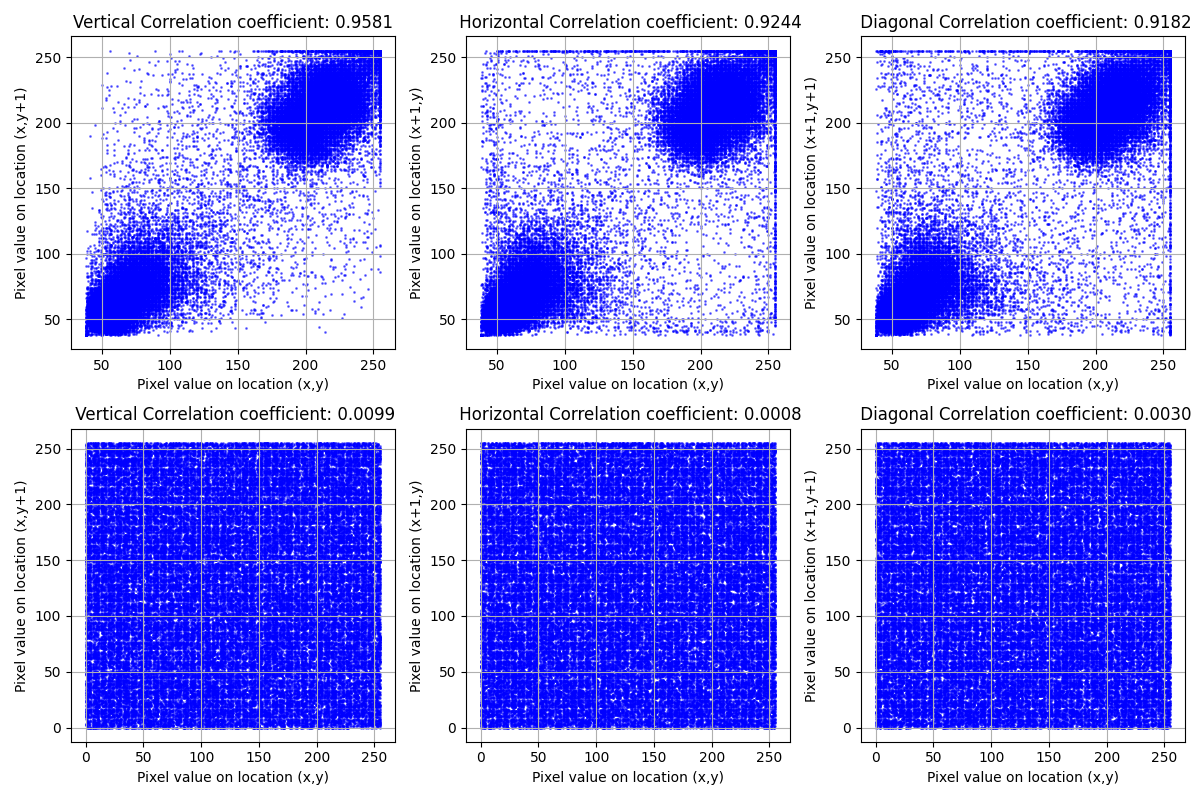}
         \caption{$LCT_2$}
     \end{subfigure}
    
        \caption{Pixel Correlation Analysis}
\end{figure}
The correlation coefficients for all 3 directions are presented in Table 5, demonstrating the statistical attack resistance of the proposed encryption scheme on the selected sample medical images.

 \begin{table}[h!]
\caption{Correlation Analysis}
\label{tab1e}
\centering
\begin{tabular}{lcc|cc|cc}
\hline
& \multicolumn{2}{c}{\textbf{Horizontal Correlation}} & \multicolumn{2}{c}{\textbf{Vertical Correlation}} & \multicolumn{2}{c}{\textbf{Diagonal Correlation}} \\ [0.5ex]
\hline 
 \textbf{Image}  & Original & Encrypted & Original & Encrypted & Original & Encrypted \\
\hline
$BMRI_1$ & 0.9596 & 0.0016 & 0.9605 & 0.0016 & 0.9214 & 0.0044\\
$BMRI_2$ & 0.9540 & 0.0036 & 0.9593 & 0.0074 & 0.9787 & -0.1184\\
$CXR_1$ & 0.9929 & -0.0002 & 0.9929 & -0.0088 & 0.9973 & -0.0776\\
$CXR_2$ & 0.9957 & -0.0144 & 0.9932 & 0.0007 & 0.9905 & 0.0039\\
$LCT_1$ & 0.9835 & -0.0040 & 0.9576 & -0.0005 & 0.9466 & 0.0045\\
$LCT_2$ & 0.7084 & 0.0002 & 0.9553 & 0.0107 & 0.9482 & -0.0075\\
\hline
\end{tabular}
\end{table}

\subsection{Differential Attack Analysis}

The sensitivity of the encryption technique to even minor changes in the plain image is evaluated using the differential attack. The Number of Pixel Change Rate (NPCR) and Unified Average Change in Intensity (UACI) are the two most important performance metrics to assess the proposed technique's resistance to differential attacks \cite{Khan2022ImageNetwork}. NPCR decides the pixel rate in the encrypted images whenever a single pixel of the test image is changed. It is employed to evaluate the resistance to differential attack. Higher than 99\% is the ideal value of NPCR. NPCR is calculated using the following Equations 18 and 19. 
\begin{equation}
    NPCR = \frac{\Sigma_{p,q}DF(p,q)}{\mathcal{W} \times \mathcal{H}} \times 100\%
\end{equation}
Here,
\begin{equation}
    DF(p,q)=\left\{
    \begin{array}{ll}
      1, & \mbox{if $Med(p,q)$ equals $CMed(p,q)$}\\
      0, & \mbox{if $Med(p,q)$ not equals $CMed(p,q)$}
    \end{array}
  \right.
\end{equation}
where $\mathcal{W}$ and $\mathcal{H}$ denote the width and height of the image. $DF(p,q)$ is the function that calculates the difference between the respective pixels of the original medical image $Med$ and the encrypted medical image $CMed$.

 \begin{table}[h!]
\caption{Differential Attack Analysis}
\label{tab1e}
\centering
\begin{tabular}{lcc}
\hline
\textbf{Image} & \textbf{NPCR} & \textbf{UACI} \\ [0.5ex]
\hline
$BMRI_1$ & 99.59 & 33.61 \\
$BMRI_2$ & 99.63 & 33.69 \\
$CXR_1$ & 99.61 & 33.01 \\
$CXR_2$ & 99.54 & 33.07 \\
$LCT_1$ & 99.60 & 34.33 \\
$LCT_2$ & 99.61 & 33.86 \\
\hline
\end{tabular}
\end{table}

The Unified Average Changing Intensity (UACI), quantifies the average disparity in pixel intensity between the original and encrypted images. This metric is frequently utilized to indicate resilience against a differential attack. An optimal UACI value hovers around 33\%, calculated according to Equation 20.

\begin{equation}
    UACI = \frac{\Sigma_{p,q} Med(p,q) - CMed(p,q)}{255 \times \mathcal{W}\times \mathcal{H} } \times 100\%
\end{equation}
Table 6 shows the NPCR and UACI of the sample medical images and proves the withstanding power of the proposed encryption algorithm.
\subsection{Entropy Analysis}
Entropy characterizes the unpredictability of image information, denoted by $Ey$. It serves to measure the uncertainty inherent in the proposed encryption technique and is computed using Equation 21.

\begin{equation}
     Ey = - \Sigma_{k=1}^{255} P_k log(P_k)
\end{equation}
The probability of occurrence of pixel value $k$ is denoted by $P_k$. The value of $Ey$ \begin{math}
    \in [0,8]
\end{math}. For an 8-bit image, the entropy value should approach 8. Table 7 illustrates the entropy values of the chosen sample images, revealing that the entropy of all encrypted images closely approximates 8.

 \begin{table}[h!]
\caption{Entropy Analysis}
\label{tab1e}
\centering
\begin{tabular}{lcc}
\hline
& \multicolumn{2}{c}{\textbf{Entropy}} \\ [0.5ex]
\hline 
 \textbf{Image}  & Original & Encrypted  \\
\hline
$BMRI_1$ & 6.9816 & 7.9971 \\
$BMRI_2$ & 6.9592 & 7.9969 \\
$CXR_1$ & 7.6995 & 7.9969 \\
$CXR_2$ & 7.6933 & 7.9970 \\
$LCT_1$ & 6.2786 & 7.9970 \\
$LCT_2$ & 6.9756 & 7.9969 \\
\hline
\end{tabular}
\end{table}

\subsection{Error Metrics}
There are few metrics available to assess the error of the encryption model. MAE, RMSE, and PSNR are standard metrics to assess whether the encryption scheme produces acceptable errors. MAE is used to measure the variance between encrypted and original images. MAE \begin{math} \in [0,2^m-1] \end{math}, where m is the number of bits to represent each pixel. For a good encryption model, MAE should be maximum. It can be evaluated using Equation 22.

\begin{equation}
    MAE = \frac{1}{\mathcal{W}\times \mathcal{H}}\Sigma_{p,q} |CMed(p,q) - Med(p,q)|
\end{equation}

MSE is useful for comparing exact pixel values between an original image and an encrypted image. The error is the difference between the original image's and the encrypted image's pixel values. In order to provide more precise and reliable data, RMSE assesses the MSE root. A desirable encryption algorithm would yield a minimal RMSE value. Equations 23 and 24 can be used to calculate these measures.

\begin{equation}
    MSE = \frac{1}{\mathcal{W}\times \mathcal{H}}\Sigma_{p,q} (CMed(p,q) - Med(p,q))^2
\end{equation}
\begin{equation}
    RMSE = \sqrt{MSE}
\end{equation}

The range of RMSE \begin{math} \in [0,\infty] \end{math}.
PSNR is used as a quality measurement between the original and decrypted images. PSNR is mathematically computed as follows in Equation 25.
\begin{equation}
    PSNR = 10 log_{10} \frac{(2^m-1)^2}{MSE}
\end{equation}
where $n$ represents the number of bits per pixel. PSNR is measured in decibels (dB). Table 8 shows the error metrics of the sample medical images in the proposed model.

\begin{table}[h!]
\caption{Error Metrics}
\label{tab1e}
\centering
\begin{tabular}{lccc}
\hline
\textbf{Image} & \textbf{MAE} & \textbf{RMSE} & \textbf{PSNR(dB)}\\ [0.5ex]
\hline
$BMRI_1$ & 127.24 & 104.97 & 7.71 \\
$BMRI_2$ & 127.46 & 105.07 & 7.70 \\
$CXR_1$ & 127.33 & 103.13 & 7.86 \\
$CXR_2$ & 127.98 & 103.28 & 7.85 \\
$LCT_1$ & 127.30 & 107.14 & 7.53 \\
$LCT_2$ & 127.62 & 105.72 & 7.65 \\
\hline
\end{tabular}
\end{table}

\subsection{Known Plaintext Attack Analysis}

In the realm of image encryption, these assaults pose challenges, especially given the substantial data content typically found in images, making them complex targets for protection. Through these methods, adversaries aim to anticipate the link between plaintext and ciphertext images, seeking recurring patterns to deduce the secret encryption key or restore the original image. To counter known-plaintext attacks in image encryption, various plaintext images such as "Black" and "White" are selected as test subjects for encryption using the proposed QMedShield, as depicted in Figure 15, with the aim of undermining the scheme's integrity.

\begin{figure}[h!]
     \centering
     \begin{subfigure}[b]{0.4\textwidth}
         \centering
         \includegraphics[width=0.7\textwidth]{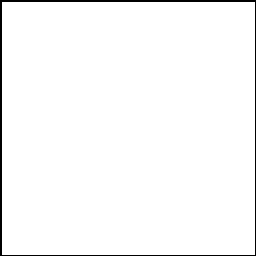}
         \caption{White Image}
     \end{subfigure}
     \hfill
     \begin{subfigure}[b]{0.4\textwidth}
         \centering
         \includegraphics[width=0.7\textwidth]{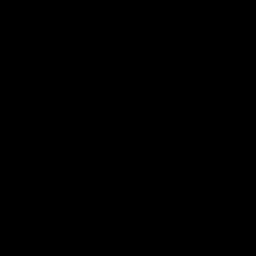}
         \caption{Black Image}
     \end{subfigure}
     \hfill
     \begin{subfigure}[b]{0.4\textwidth}
         \centering
         \includegraphics[width=0.7\textwidth]{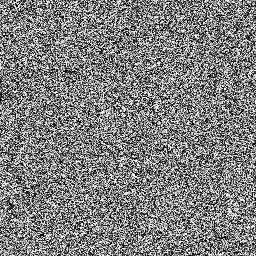}
         \caption{Encrypted White Image}
     \end{subfigure}
      \begin{subfigure}[b]{0.4\textwidth}
         \centering
         \includegraphics[width=0.7\textwidth]{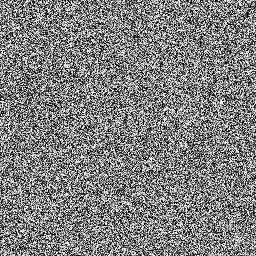}
         \caption{Encrypted Black Image}
     \end{subfigure}
        \caption{KP Attack Simulation}
\end{figure}

\subsection{Chosen Plaintext Attack Analysis}
Given that attackers exert greater influence over plaintext images, the chosen-plaintext attack holds more potency than the known-plaintext attack. Consequently, an image encryption algorithm capable of thwarting a chosen-plaintext attack is inherently resilient against a known-plaintext attack as well. To disperse pixel values effectively, QMedShield employs an XOR operation, while Equation 26 is utilized to assess its resistance against chosen-plaintext attacks.

\begin{equation}
    Med_1(p,q)\oplus Med_2(p,q) = CMed_1(p,q) \oplus CMed_2(p,q)
\end{equation}

\begin{figure}[h!]
     \centering
     \begin{subfigure}[b]{0.4\textwidth}
         \centering
         \includegraphics[width=0.5\textwidth]{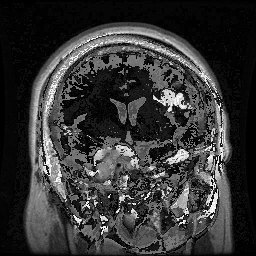}
         \caption{$Med_1(p,q)\oplus Med_2(p,q)$}
     \end{subfigure}
     \hfill
     \begin{subfigure}[b]{0.4\textwidth}
         \centering
         \includegraphics[width=0.5\textwidth]{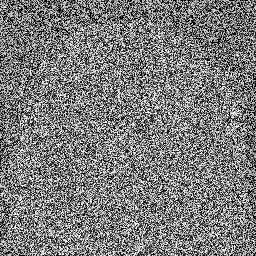}
         \caption{$CMed_1(p,q) \oplus CMed_2(p,q)$}
     \end{subfigure}
     \caption{CP Attack Simulation}
\end{figure}

In this scenario, $Med_1$ and $Med_2$ represent 2 test images and their respective encrypted cipher images are $CMed_1$ and $CMed_2$.  If equation 26 is fulfilled, then the CP attack is possible. The proposed method offers protection against such attacks, as illustrated in Figure 16, where equation 26 is found to be unjustified.

\subsection{Comparative Security Analysis}

Proposed QMedShield is compared with other existing image encryption models \cite{Liu2017ASequence}\cite{Janani2021ARepresentation}\cite{AmaithiRajan2023SecureEncoding}\cite{Patel2024SecuredMethods}. We have taken correlation coefficients, NPCR, UACI, and Entropy metrics to compare our proposed method with existing encryption models. Table 9 shows the better correlations and NPCR of the proposed model. Thus the proposed QMedShield contains quantum chaotic maps, DNA-encoding, and bit plane scrambling yields an improved UACI value, and reduces the correlation coefficient among the pixels in the encrypted image. An improved UACI value indicates that the QMedShield is effectively dispersing pixel values, which is desirable for enhancing image security.  A reduction in correlation coefficients shows that the QMedShield is disrupting any patterns or relationships between pixel values better, making it more challenging for attackers to decipher the original image. Figures 17 (a) and (b) depict the comparative analysis visually. This clearly shows that the proposed algorithm is better than the existing encryption techniques. 

\begin{table}[h!]
\caption{Encryption Model: Comparative Analysis}
\label{tab1e}
\centering
\begin{tabular}{lllllll}
\hline
\textbf{Reference} & \textbf{HCorr} & \textbf{VCorr} & \textbf{Dcorr} & \textbf{NPCR} & \textbf{UACI} & \textbf{Entropy}\\ [0.5ex]
\hline
Liu et al.\cite{Liu2017ASequence}& 0.0002 & 0.0002 & 0.0005 & 99.68 & 33.44 & -\\
Janani et al.\cite{Janani2021ARepresentation} & -0.0045 & 0.0103 & 0.0011 & 99.72 & 33.46 & -\\
Amaithi Rajan et al.\cite{AmaithiRajan2023SecureEncoding}& -0.0011 & -0.0154 & -0.0282 & 99.61 & 33.39 & 7.9971\\
Patel et al.\cite{Patel2024SecuredMethods}& 0.0020 & -0.0012 & 0.0043 & 99.75 & 33.51 & 7.9975\\
QMedShield (Ours) & \textbf{-0.0347} & \textbf{-0.0211} &\textbf{-0.0418} & \textbf{99.62} & \textbf{33.58} & \textbf{7.9971}\\
\hline
\end{tabular}
\end{table}

\begin{figure}[h!]
     \centering
     \begin{subfigure}[b]{0.45\textwidth}
         \centering
         \includegraphics[width=\textwidth]{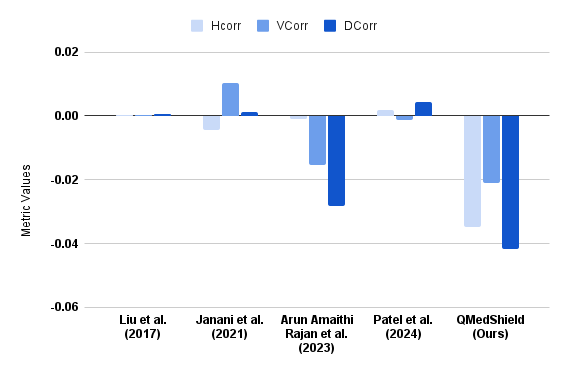}
         \caption{Correlation Co-efficients}
     \end{subfigure}
     \hfill
     \begin{subfigure}[b]{0.45\textwidth}
         \centering
         \includegraphics[width=\textwidth]{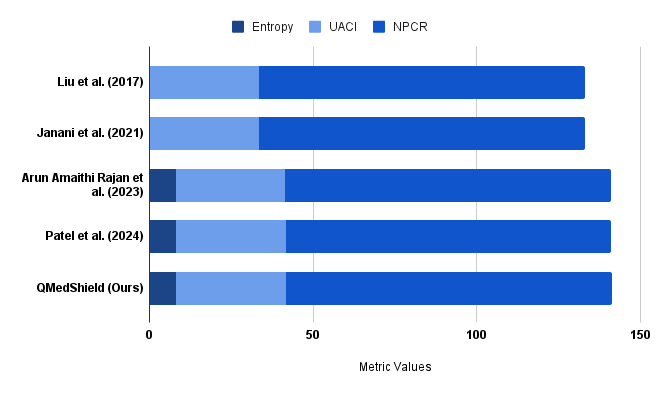}
         \caption{Entropy, NPCR, UACI values}
     \end{subfigure}
     \caption{Comparative Analysis}
\end{figure}

\section{Conclusion and Future works}

In light of the quantum computing era's emergence, conventional security protocols are increasingly vulnerable. To address this challenge, this paper introduces QMedShield, a novel encryption algorithm tailored for medical images. QMedShield combines bit-plane scrambling with quantum chaotic maps for pixel diffusion, alongside hybrid-chaotic maps and DNA-encoding techniques for pixel confusion. Extensive testing on three diverse medical datasets, coupled with rigorous statistical analyses and attack-resistant evaluations, demonstrates QMedShield's robustness against numerous threats. Looking ahead, future advancements could explore the application of high-dimensional quantum chaotic maps to further enhance pixel diffusion. Additionally, optimizing the storage of image pixels through efficient qubit-based pixel substitution holds promise. Moreover, the integration of scrambling techniques for both row and column pixels across each bit-plane could bolster the diffusion rate, contributing to the algorithm's overall security and effectiveness in safeguarding sensitive medical imagery.



\end{document}